\documentclass[review,12pt]{elsarticle}

\biboptions{square,comma,sort&compress}

\usepackage{amssymb} % The amssymb package provides various useful mathematical symbols
\usepackage{amsthm} % The amsthm package provides extended theorem environments
\usepackage{amsmath}
\usepackage{mathrsfs}
\usepackage{graphicx}
\usepackage{epstopdf}
\usepackage{float}
\usepackage{caption}
\usepackage{soul}
\usepackage{color,soul} %to highlight text
\usepackage{color} %To highlight references in the text
\usepackage{subcaption}
\usepackage{bm}
\usepackage{bbm}
\usepackage{mathrsfs}
\usepackage{soul}
\usepackage[margin=2cm]{geometry}% to reduce margins
\usepackage{booktabs}
\usepackage{chngpage}
\usepackage{enumitem}%enumeration
\usepackage{subcaption}
\usepackage{multirow}
\usepackage{stackengine} %overlap plots
\usepackage{mhchem} %chemical formulas
\usepackage{hyperref} %links
\usepackage{cleveref}
\usepackage{upgreek}
\usepackage{textcomp} 

\usepackage{lineno}
\journal{Nuclear Materials and Energy}

\makeatletter
\def\@author#1{\g@addto@macro\elsauthors{\normalsize%
    \def\baselinestretch{1}%
    \upshape\authorsep#1\unskip\textsuperscript{%
      \ifx\@fnmark\@empty\else\unskip\sep\@fnmark\let\sep=,\fi
      \ifx\@corref\@empty\else\unskip\sep\@corref\let\sep=,\fi
      }%
    \def\authorsep{\unskip,\space}%
    \global\let\@fnmark\@empty
    \global\let\@corref\@empty  %% Added
    \global\let\sep\@empty}%
    \@eadauthor={#1}
}
\makeatother

\makeatletter
\def\ps@pprintTitle{%
  \let\@oddhead\@empty
  \let\@evenhead\@empty
  \def\@oddfoot{\reset@font\hfil\thepage\hfil}
  \let\@evenfoot\@oddfoot
}
\makeatother

\makeatletter
\def\thickhline{%
  \noalign{\ifnum0=`}\fi\hrule \@height \thickarrayrulewidth \futurelet
   \reserved@a\@xthickhline}
\def\@xthickhline{\ifx\reserved@a\thickhline
               \vskip\doublerulesep
               \vskip-\thickarrayrulewidth
             \fi
      \ifnum0=`{\fi}}
\makeatother

\newlength{\thickarrayrulewidth}
\begin{document}

\begin{frontmatter}



\title{Understanding the oxidation of pure Tungsten in air and its impact on the lifecycle of a fusion power plant}


\author{Rongrui Li\fnref{IC,OX}}

\author{Guillermo \'{A}lvarez\fnref{IC,UO}}

\author{Ayla Ipakchi\fnref{IC}}

\author{Livia Cupertino-Malheiros\fnref{IC}}

\author{Mark R. Gilbert\fnref{UKAEA,IC2}}

\author{Emilio Mart\'{\i}nez-Pa\~neda\corref{cor1}\fnref{OX}}
\ead{emilio.martinez-paneda@eng.ox.ac.uk}

\author{Eric Prestat\corref{cor1}\fnref{UKAEA,IC}}
\ead{eric.prestat@ukaea.uk}

\address[IC]{Department of Civil and Environmental Engineering, Imperial College London, London SW7 2AZ, UK}

\address[OX]{Department of Engineering Science, University of Oxford, Oxford OX1 3PJ, UK}

\address[UO]{Department of construction and manufacturing engineering, University of Oviedo, Polytechnic School of Engineering of Gijón, East Building, 33203, Asturias, Spain}

\address[UKAEA]{United Kingdom Atomic Energy Authority, Culham Campus, Abingdon, OX14 3DB, UK}

\address[IC2]{Department of Mechanical Engineering,  Imperial College London, London SW7 2BX, UK}
\cortext[cor1]{Corresponding author.}

\begin{abstract}

\noindent The oxidation of pure W and the sublimation of W oxide have been investigated to assess their impact on the lifecycle of a fusion power plant. Pure W has been oxidised at temperatures between 400 and 1050°C and for durations ranging between 1 and 70 h. The formation of voids and cracks has been observed at temperatures above 600°C, leading to the formation of dust or oxide spalling, which could be problematic in maintenance and waste-handling scenarios of a fusion power plant. Preferential oxidation taking place at the edge of the specimen was characterised, and its impact is discussed in relation to component design. Characterisation using electron microscopy and Raman spectroscopy revealed that the oxide scale is formed of three main layers:  the inner layer is 30-50 nm thick WO$_2$ oxide, the middle layer is a 10-20 \textmu m of WO$_{2.72}$ and the outer layer is formed of  WO$_{2.9}$/WO$_3$ phases – whose thickness varies according to the total thickness of the oxide scale. The observed microstructure is discussed in relation to the parabolic-to-linear kinetics and its potential impact on tritium permeation and detritiation efficiency.\\

\end{abstract}



\end{frontmatter}
%\linenumbers

%\begin{figure}[H]
%     \centering
%         \centering
%         \includegraphics[width=1\textwidth]{GraphicalAbstract.jpg}
%        \label{fig:GraphicalAbstract}
%\end{figure}
\section{Introduction}
\label{Introduction}

Tungsten (W) has emerged as the favoured choice for the divertor and first wall plasma-facing material (PFM) due to its high melting point (3422°C), excellent thermal conductivity, low sputtering erosion, and minimal fuel retention \cite{tokitani2019demonstration,koch2007self,zhang2023microstructure}. However, W oxidises readily with a very low partial pressure of oxygen \cite{HABAINY201826}. This can result in significant loss of first wall materials and dispersion of W oxide in the form of sublimated gas molecules \cite{klein_oxidation_2018,klein_studies_2020,nagy2022oxidation} or dust spalled off from the oxide scale. This can have an impact on the operability and the sustainability of the fusion device. Oxidation is expected to be a concern when the W components are at a temperature higher than 500°C and exposed to air, which can occur in maintenance, accident or waste handling scenarios. Indeed, neutron irradiation during normal operation will cause high gamma activity and heat output from irradiated W in plasma-facing components. Stopping the active cooling will then cause the temperature of W to increase significantly \cite{noce_2021_NeutronicsAnalysisActivation,terentyev_2021_NeutronIrradiationHardening}. This will happen in the following scenarios: accident scenario (loss of coolant accident (LOCA) combined with loss of vacuum accident (LOVA)); maintenance scenario, where the inner vessel will be exposed to air; waste handling and detritiation scenario. It has been reported that the detritiation efficiency in W components from the Joint European Torus (JET) is low and that the oxide scale may be acting as a permeation barrier \cite{stokes_detritiation_2023}. Recent modelling results predict that one of the W oxide phases, WO$_2$, could act as a tritium barrier \cite{christensen_atomistic_2024}, explaining the previously reported low detritiation efficiency. Therefore, it is important to understand the oxidation behaviour and what is the microstructure of the W oxide scale (oxide phase formed and morphology), as this will define the stability of the oxide scale and inform the impact of dust formation through oxide spalling, as well as potentially allowing for the optimisation of the detritiation of W components. 

The interaction of oxygen with metal substrates involves three primary processes: adsorption, dissolution, and diffusion \cite{lassner1999tungsten,gupta2003thermodynamics}.
Diffusion acts as the rate-determining step (RDS) that collectively determines the overall rate of oxidation \cite{hauffe1995mechanism}. It is particularly influential in the case of protective metal oxide layers via point defect mechanisms \cite{macdonald_1992_PointDefect}. Marker movement observed using the interruption kinetics technique indicates that oxygen anion vacancies are the primary defects responsible for ion transport \cite{sarrazin1996contribution,sikka1980oxidation}. Wagner's theory was proposed to quantitatively describe oxide scale formation, suggesting the oxidation rate depends on the electrochemical potential difference at the metal-oxide and oxide-gas interfaces \cite{landolt2007corrosion}. This theory posits that the oxide layer's growth rate is governed by the migration of metal cations and oxygen anions through the oxide layer, with the layer initially forming at the metal-oxygen boundary and growing due to ion transport. The transport number in high-temperature solid-state diffusion represents the fraction of ionic current carried by cations or anions \cite{kofstad1995defects}. In W trioxide (WO$_{3}$), a group VIb binary oxide, the anion transport number is 0.63, higher than the cation's 0.37 \cite{landolt2007corrosion}. This indicates that oxygen anions play a dominant role in the ionic diffusion flux during WO$_{3}$ formation. 

The oxidation of W typically proceeds through three kinetic regimes, with the RDS shifting depending on temperature and oxide scale thickness: At a temperature below 400$^\circ$C, oxidation proceeds via a logarithmic law \cite{Birks2012}. As the inner compact oxide layer thickens with time, the barriers to oxygen become more noticeable. The reaction then shifts to diffusion control according to Wagner theory \cite{samal2016high‐temperature}. The inward diffusion of oxygen ions becomes the RDS, rather than the interfacial reaction at the metal–metal oxide (M:MO) interface, leading to the characteristic parabolic growth kinetics \cite{sarrazin1996contribution,sikka1980oxidation}. At higher temperatures or after prolonged exposure, tungsten oxidation transitions into a linear growth regime. This behaviour has frequently been attributed to microcracking, which causes the lost of the diffusion barrier protection \cite{gulbransen1960kinetics,CIFUENTES2012114}. Previous investigations into tungsten oxidation have consistently identified both parabolic and linear oxidation regimes, with the formation of WO$_3$ as the predominant oxide phase \cite{gulbransen1960kinetics,nagy2022oxidation}. However, for the application of tungsten in fusion energy systems, further insights are required in two critical areas:
\begin{itemize}
    \item Spalling behaviour during extended oxidation durations (beyond the ~10 hours typically examined), due to its significant role in dust generation in accident (combined LOCA LOVA), maintenance, waste handling or detritiation scenario.
    \item The microstructural characteristics of the oxide scale, which are expected to influence detritiation efficiency and hydrogen isotope retention.
\end{itemize}

To address these knowledge gaps, the present study undertakes long-duration oxidation experiments and provides a detailed characterisation of the resulting oxide scale microstructure.

\section{Materials and experimental methods}
\label{Sec:Experimental}

\subsection{Materials}
\label{Subsec:Materials}
In this work, pure W (99.95\%) supplied by Future Alloys LTD and manufactured using forging was used to analyse the oxidation of the material in dry air. It was cut by electrical discharge machining (EDM) from a W rod and ground to a P2000 grit. W oxides powders were sourced to be used as characterisation standards. WO$_{2}$ and WO$_{3}$ powders were supplied by Merck Life Science UK Ltd (part number 400505 and 95410). Optical microscopic inspection of the as-received WO$_{3}$ powder revealed sporadic dark-blue particles, indicating the presence of oxygen-deficient W oxides. Such sub-stoichiometric oxides, belonging to the W$_{n}$O$_{3n-1}$ or W$_{n}$O$_{3n-2}$ series \cite{mardare_2019_ReviewVersatility}, are commonly formed during the manufacturing process, despite the nominal purity of the WO$_3$ powder being 99.9\%. Given the small fraction of sub-stoichiometric oxides, the powder can be reasonably considered to be predominantly monoclinic WO$_3$ at room temperature~\cite{lunk2023molybdenum}. WO$_{2.72}$ (W$_{18}$O$_{49}$) was supplied by Stanford Advanced Materials (US) and WO$_{2.9}$ (W$_{20}$O$_{58}$) was supplied by Thermo Fisher Scientific Ltd.

\begin{figure}[H]
     \centering
    \begin{subfigure}{0.7\textwidth}
         \centering
         \includegraphics[width=\textwidth]{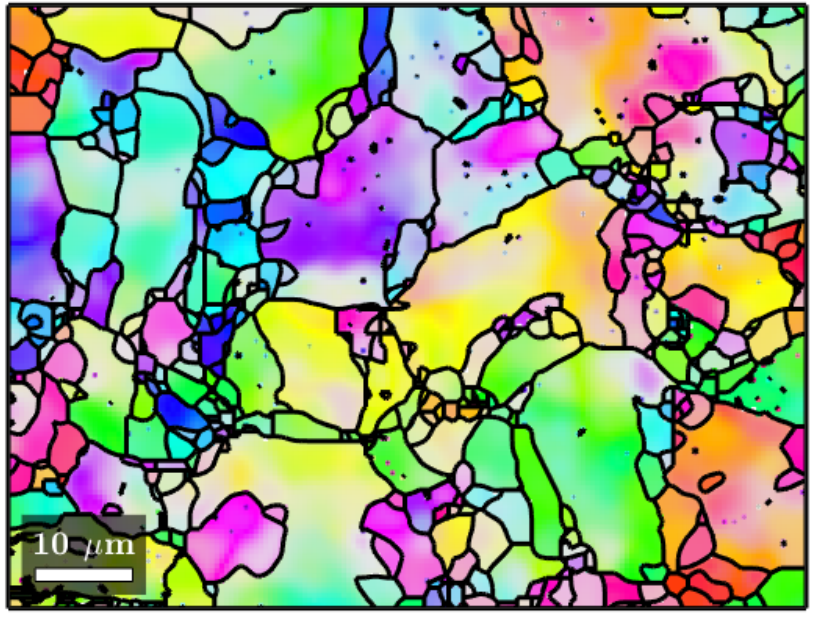}
         \end{subfigure}\hspace{1cm}
    \begin{subfigure}{0.2\textwidth}
         \centering
         \includegraphics[width=\textwidth]{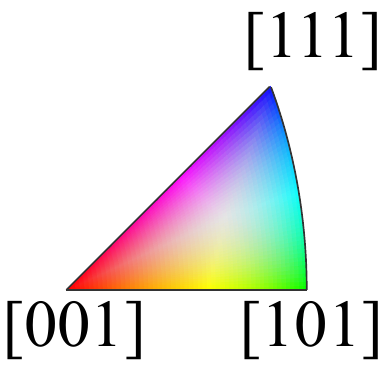}
         \end{subfigure}\hfill
        \caption{EBSD inverse pole figure of pure W before oxidation. The grain boundaries are highlighted in black, with a misorientation of $>5^{\circ}$.}
        \label{fig:EBSD1}
\end{figure}

The inverse pole figure map (Fig. \ref{fig:EBSD1}) obtained by EBSD analysis shows that the microstructure of pure W consists of a random distribution of grain sizes and orientations.

\subsection{Simultaneous thermal analysis}
\label{Subsec:STA}

Simultaneous thermal analysis (STA) combines thermogravimetric analysis (TGA) and differential scanning calorimetry (DSC). A NETZSCH STA 449 F3 Jupiter was used to analyse the oxidation of pure W and sublimation of W trioxide (WO$_3$). Temperature calibration of the STA was performed using standard reference metals (In, Sn, Zn, Al, Au) with known melting points.

For the experiments in dry air, the specimen was placed in the hanging holder at room temperature. Once the furnace is closed and insulated from the external ambient environment, three evacuation and backfill cycles were applied to reduce oxygen impurities during the heat ramp. The reduction of the residual oxygen inside the furnace, after three purges, was confirmed with a mass spectrometer connected to the furnace. After this purging process, the temperature was increased until it reached the desired one using a heating rate of 10 K/min under a nitrogen atmosphere. Before introducing the synthetic air (20\% O$_2$ in N$_2$), an isothermal of 15 min in pure N$_2$ was used to stabilise the microbalance. A flow rate of 70 ml/min at 1 atm was used for all measurements. During oxidation, the partial pressure of oxygen was 0.2 atm. In all cases, the change in mass resulting from the increase in temperature before introducing synthetic air was less than 1\% of the final gain in weight.

\begin{figure}[H]
     \centering
    \begin{subfigure}{0.52\textwidth}
         \centering
         \includegraphics[width=\textwidth]{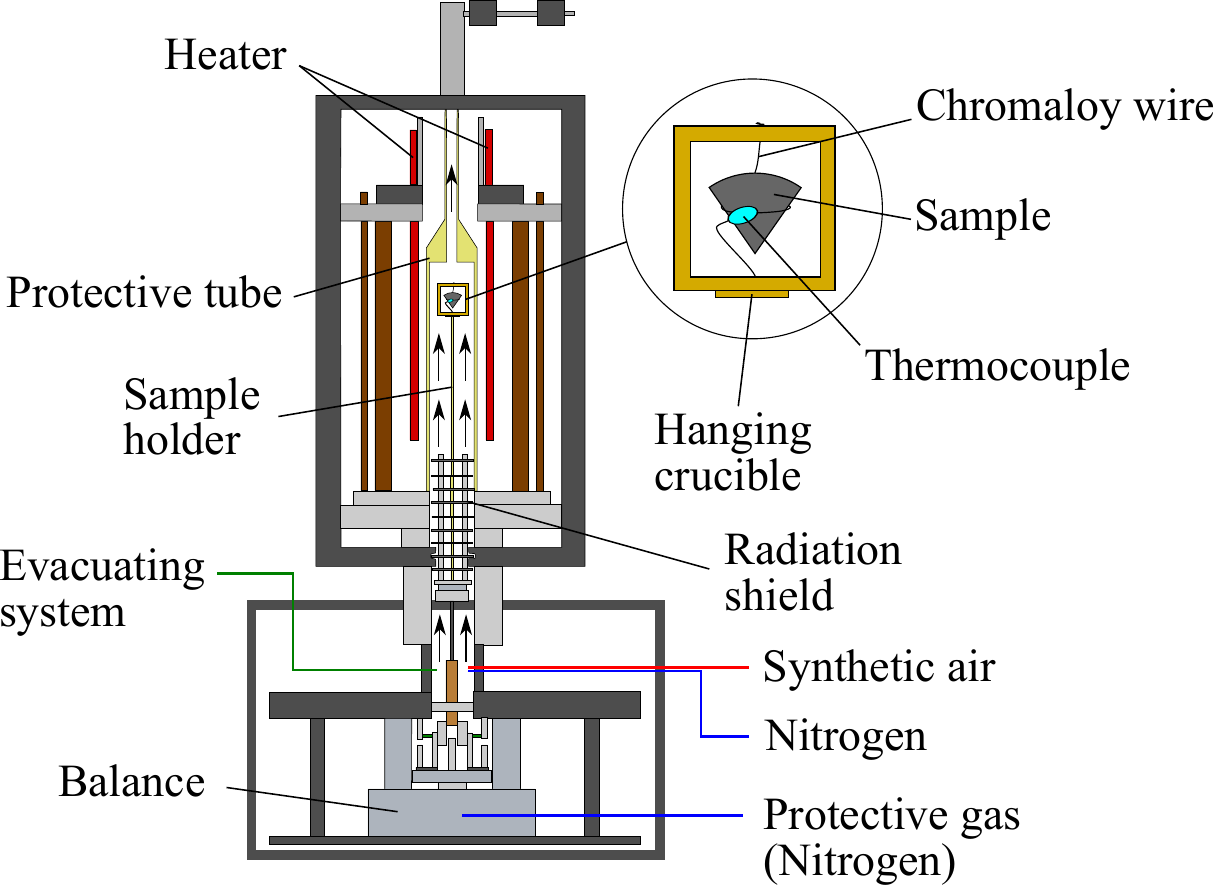}
            \caption{}
         \end{subfigure}\hfill
    \begin{subfigure}{0.47\textwidth}
         \centering
         \includegraphics[width=\textwidth]{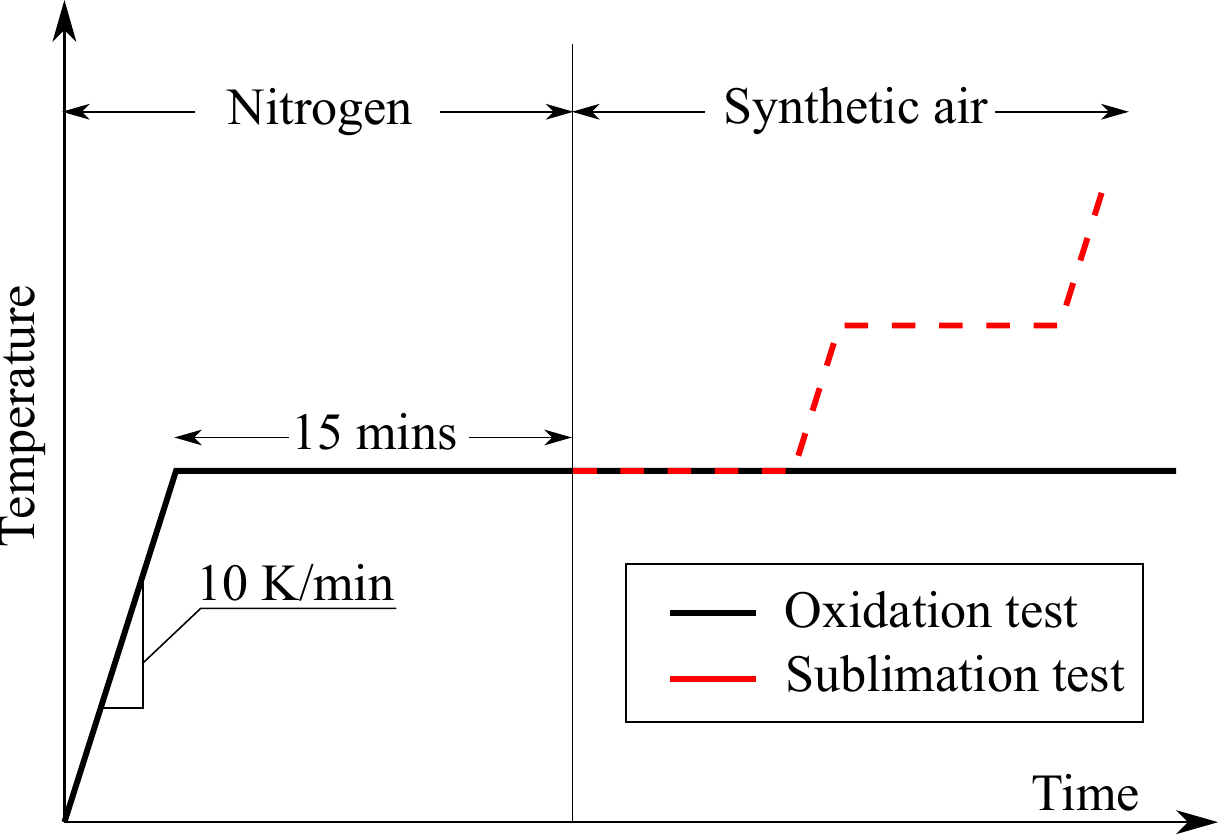}
            \caption{}
         \end{subfigure}\hfill
    \caption{(a) Schematic of the experimental setup used for oxidation. The specimen is held with a Chromaloy wire using the hanging holder. (b) Program used for oxidation and sublimation measurements showing the heat ramps, the isothermals and the environment changing from pure N$_2$ to synthetic air.}
    \label{fig:experimental_setup}
\end{figure}

For the oxidation process, the TGA technique was used to record the mass gain in the sample as a function of time. The standard specimen shapes (unless specified otherwise) were a quarter of a circle with nominal diameter and thickness values of 20 mm and 0.5 mm, respectively (approximately 800 mg). In the first stage, different sample holders were tested to identify the most suitable configuration for pure W oxidation. The hanging holder was selected because it offers better exposure of the sample surface to the ambient atmosphere, resulting in more uniform oxidation compared to the plane and cone holders. To hang the specimen, a Chromaloy\textsuperscript{\textregistered} O-Resistance wire (0.5 mm in diameter) was used due to its excellent high-temperature oxidation resistance and thermal stability behaviour, as reported by Habainy et al. \cite{HABAINY201826}. Fig. \ref{fig:experimental_setup} illustrates the configuration of the hanging sample carrier for a pure W sample. The temperature was measured using a thermocouple located within 5 mm of the specimens.

For the sublimation process, the combination of TGA and DSC was used to capture the thermal and mass responses of the W oxide. Pure WO$_{3}$ powder, described in Section \ref{Subsec:Materials}, was used to avoid the influence of oxidation on the metal and characterise only WO$_{3}$ sublimation.
The DSC technique was selected in order to detect the different phase changes produced at different temperatures under synthetic air. As in the oxidation tests, three purges were performed in the same manner. Then, a temperature ramp between RT and the selected initial temperature was programmed (10 K/min). 
The experimental procedure employed in this test involved incrementally raising the temperature, followed by maintaining a constant temperature for 2 hours to calculate the sublimation rate. In this experiment, the mass gain and heat flow were recorded over time. 

The thickness of the oxide layer has been estimated assuming that the oxide scale is composed of WO$_3$ (WO$_3$, WO$_{2.72}$ and WO$_{2.9}$ have similar densities, ranging between 7.16 and 7.78 $\mathrm{g}\cdot\mathrm{cm}^{-3}$ \cite{lassner1999tungsten}) and that added mass is associated with the intake of oxygen from the gas phase. Similarly as the approach used in \cite{SCHLUETER2019102}, this gives the following equation:
\begin{equation}
\label{eqn:thickness}
    L_{thickness} = \frac{\Delta m}{ \frac{3u_O}{u_W+3u_O} \cdot \rho_{WO_3}}
\end{equation}

\noindent where $L_{thickness}$ is the thickness of the oxide scale, $\Delta m$ represents the mass change per unit area, $\rho_{WO_3}$ is the density and $\frac{3u_O}{u_W+3u_O}$ is the ratio of the atomic mass of oxygen ($u_O$) and W oxide ($u_W$ is the atomic mass of W). 

The initial specimen area was used in the calculation of $\Delta m$. The specimen area will decrease slowly during oxidation and its reduction will be become significant when most of the specimen is oxidised. To avoid underestimation of the linear oxidation rate, the fitting of the mass gain curve was performed in for ranges of oxidation duration where the increase of mass gain is linear and the reduction in specimen area is negligeable.

\subsection{X-ray diffraction}
\label{Subsec:XRD}

X-ray diffraction (XRD) was used to identify the phases formed in the oxide scale. X-ray diffraction analysis was conducted on a Malvern Panalytical Empyrean XRD with a CuK$\alpha$ source ($\lambda$ $=$ 0.1542 nm) using the \\theta-2\\theta geometry between 20° and 60° and using a scanning speed of 0.01°/s. 

\subsection{Raman spectroscopy}
\label{Subsec:Raman}
Raman spectroscopy was performed with a Renishaw inVia Qontor instrument equipped with a 532 nm laser and a charged coupled device detector. The laser was focused using a Leica microscope. A 20$\times$ or 100$\times$ objective lens and a 2400 l/mm grating were employed. Raman spectra were recorded with a laser intensity of 10\% and exposure time between 0.5 and 1~s. Spikes and baselines were removed using HyperSpy \cite{hyperspy} and pybaselines \cite{pybaselines} libraries. The baselines were estimated using the adaptive smoothness penalised least squares algorithm with a smoothing parameter (``lam") equal to $2.5 \times 10^5$. The Raman maps of the WO$_{2.72}$ and WO$_3$ phases have been extracted by integrating the intensity of the 872 cm$^{-1}$ and 806 cm$^{-1}$ peaks, respectively.

\subsection{Microscopy analysis}
\label{Subsec:Microscope}

Scanning electron microscopy (SEM) was used to analyse the oxidation scale on the samples. After carrying out the oxidation tests, the samples were placed vertically and encased in epoxy resin to see the cross-section. After this process, the specimens were ground with SiC papers until the point where the scales of the edges and surfaces were visible. Finally, the specimen was polished with 0.25 \textmu m diamond paste. The final surface was analysed using a Zeiss Gemini scanning electron microscope operated at 20 kV and a Hitachi TM4000 microscope at 15 kV.

Electron Backscatter Scanning Diffraction (EBSD) was performed using a Zeiss Gemini SEM operated at 20 kV equipped with an Oxford EBSD detector to characterise the microstructure. The tilt of the specimen during EBSD analysis was 70° and the step size 0.5 \textmu m. EBSD data were analysed using the Matlab MTEX package \cite{Bachmann2010}. For EBSD analysis, the sample was ground and polished to 0.05 \textmu m using colloidal silica suspension.

\section{Results}
\label{Sec:Results}

\subsection{Oxidation}
\label{Subsec:AirOxidation}

\begin{figure}[H]
    \centering
    \begin{subfigure}{0.55\textwidth}
        \centering
        \includegraphics[width=\textwidth]{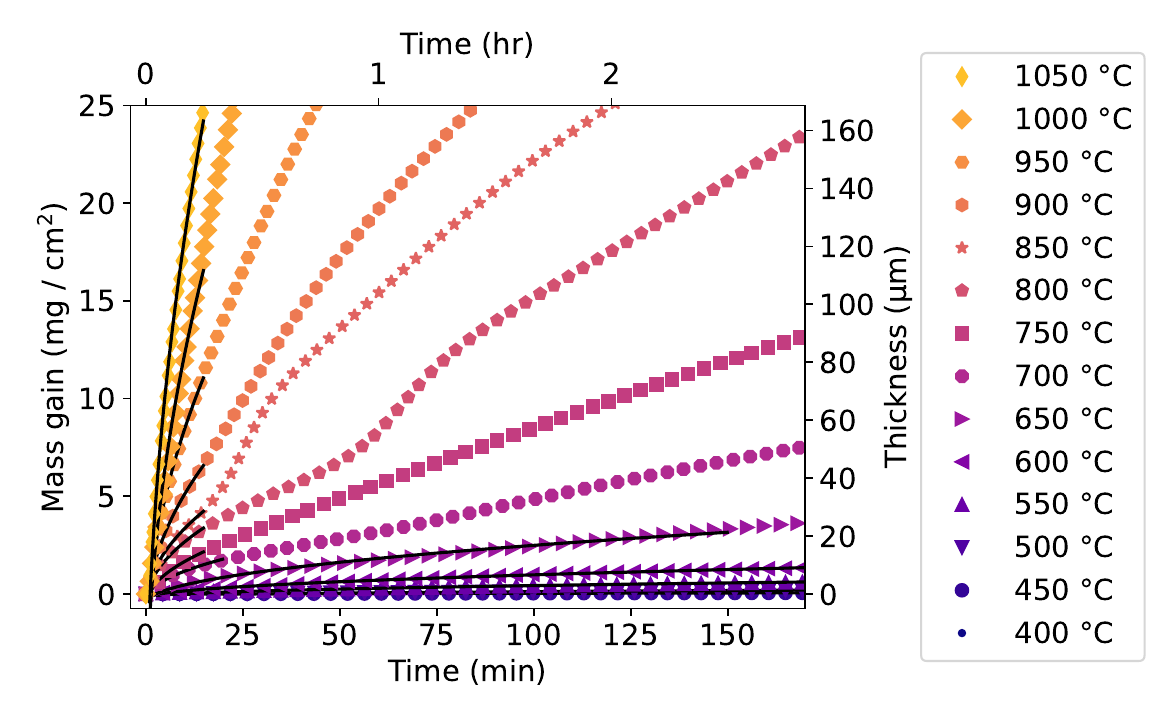}
        \caption{}
        \label{fig:TGA curves parabolic}
    \end{subfigure}\hfill
    \begin{subfigure}{0.45\textwidth}
        \centering
        \includegraphics[width=\textwidth]{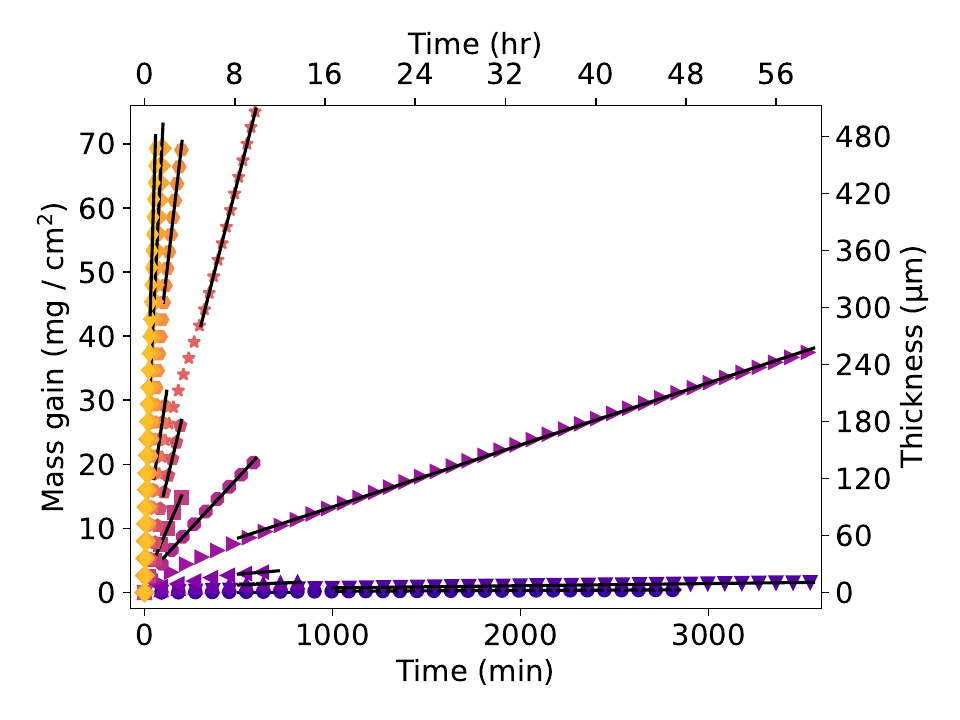}
        \caption{}
        \label{fig:TGA curves linear}
    \end{subfigure}\hfill
    \caption{Oxidation kinetics for temperatures between 400°C and 1050°C. Mass gain as a function of time for (a) the parabolic regime and (b) the linear regime. The fitted curves are plotted with solid lines.}
    \label{fig:TGA curves}
\end{figure}

Mass gains for each evaluated temperature are shown in Fig. \ref{fig:TGA curves}, where the corresponding fitted lines are plotted as black solid lines. The right-hand axis in Fig. \ref{fig:TGA curves} shows the estimated thickness of the oxide scale. The full range of oxidation is displayed in Fig. \ref{fig:TGA curves} (a), while Fig. \ref{fig:TGA curves} (b) shows the first stage of oxidation. A transition from parabolic to linear kinetics is observed, consistent with previous reports in the literature (see \cite{nagy2022oxidation} for a review).

\begin{figure}[H]
     \centering
        \includegraphics[width=0.65\textwidth]{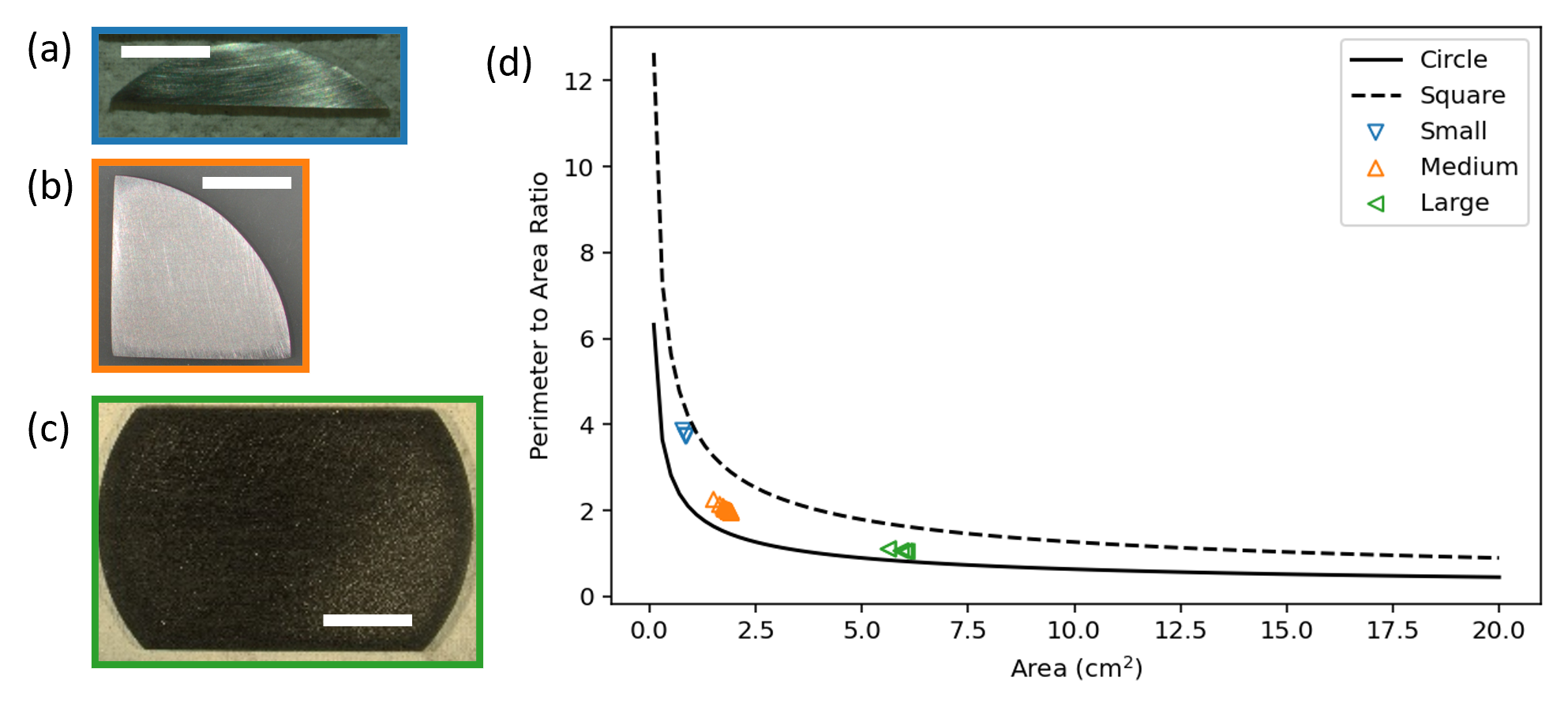}
        \caption{(a-c) The three different geometries used to assess the contribution of edges on oxidation rates. The scale bar is 5 mm. (d) The perimeter-to-area ratio (PAR) is plotted against the area and compared to a perfect geometry model (circle or square).}
    \label{fig:specimen_geometries}
\end{figure}

Various geometrical configurations have been investigated to evaluate the influence of morphological factors on oxidation rates. Particular attention was given to the preferential oxide formation at specimen edges, as compared to the bulk region where edge effects are absent. Bulk oxidation rates are expected to be the most relevant for quantification of oxidation rates in the W tiles likely to be used in fusion reactors, because the dimension of these tiles are expected to be larger (30 x 30 x 12 mm) than the test specimens used in the current study \cite{panayotis_self-castellation_2017} (i.e. the edge regions will be a relatively small component of those large tiles). To quantify the effect of edges on the measurement of the oxidation rate, we used the perimeter-to-area ratio (PAR) as a metric to measure the contribution of edges. Higher PAR means higher contribution from edges to the measurement of the oxidation rate. The ratio decays asymptotically as the size of the specimen increases, with the contribution of oxidation at the edges to the total mass gain also expected to decay in a similar way. Three different specimen geometries with varying PAR have been oxidised. The three geometries are shown in Fig. \ref{fig:specimen_geometries} and their PAR are compared to model geometries, here circle and square. Comparing the large specimen geometry - the largest that can be inserted in the TGA - with the model geometry, one can conclude that the large specimen geometry more representative of their bulk counterpart than smaller geometry since the changes in PAR of the model geometries are less than 2 for areas larger than 6 cm$^2$. For example, in the case of the square geometry, the PAR values for areas 0.8 cm$^2$ and 6 cm$^2$ are 1.8 and 4.8 times greater, respectively, than the PAR for an area of 20 cm$^2$.

\begin{figure}[H]
    \centering
    \begin{subfigure}{0.5\textwidth}
         \centering
         \includegraphics[width=\textwidth]{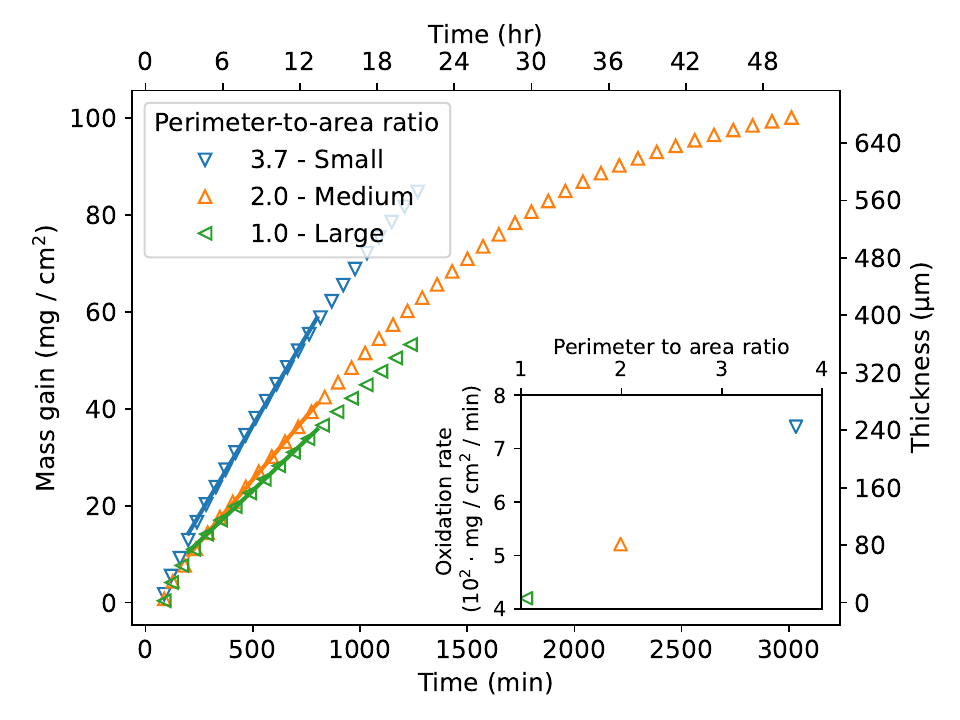}
            \caption{750°C}
         \end{subfigure}\hfill
    \begin{subfigure}{0.5\textwidth}
         \centering
         \includegraphics[width=\textwidth]{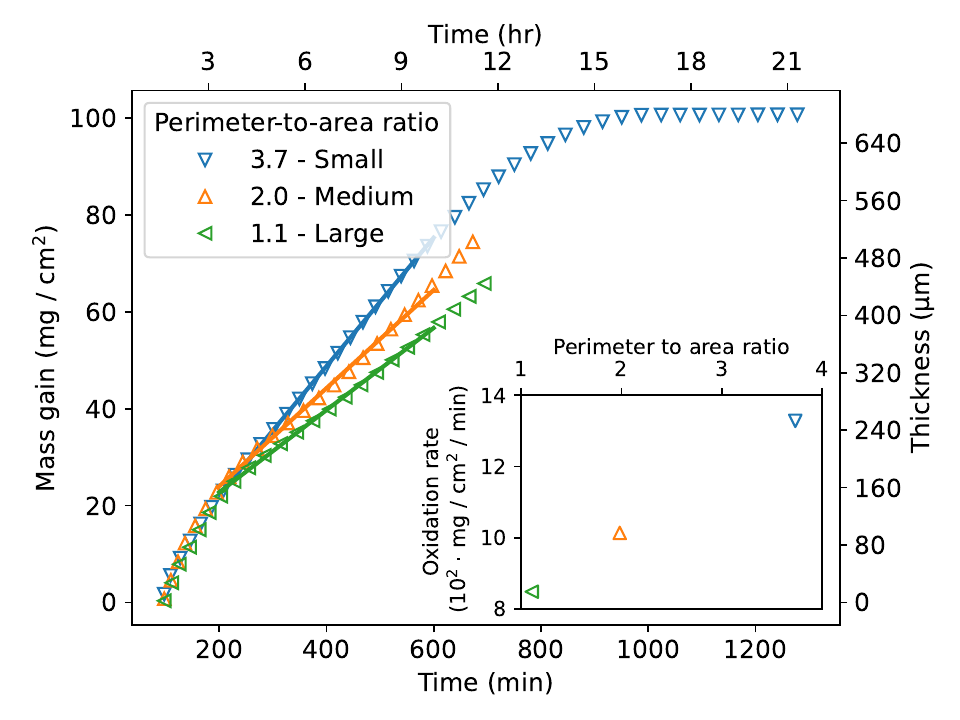}
            \caption{850°C}
         \end{subfigure}\hfill
        \caption{Mass gain as a function of oxidation time for the temperatures of (a) 750°C and (b) 850°C for the three different specimen geometries considered (small, medium and large). The dots are experimental data and the solid lines are the corresponding fits in the linear kinetics regime.}
    \label{fig:oxidation_curve_specimen_geometry}
\end{figure}

The comparison of the oxidation curve for three different specimen geometries is shown in Fig. \ref{fig:oxidation_curve_specimen_geometry}. The mass gain curves show no significant differences in the initial phase of the experiment, indicating that edge effects are negligible in the parabolic kinetic regime. This suggests that no preferential growth occurs at the edges during this stage. However, in the linear regime, following the parabolic stage, the greater PAR samples show increased oxidation rates and plotting the oxidation rate as function of the perimeter-to-area ratio (insets of Fig. \ref{fig:oxidation_curve_specimen_geometry}) shows a linear dependence. Considering that the contribution from the edge is smaller for larger geometries, this agrees with the observation of preferential oxidation taking place at the edges. At longer oxidation time (\textgreater 30 h at 750°C and \textgreater 13h at 850°C), the oxidation rate reduces due to specimen surface recession, \textit{i.e.} reduction in surface area. The small geometry specimen at 850°C is fully oxidised and therefore the mass does not change for oxidation times longer than 15 hours.

\subsection{Sublimation}
\label{Subsec:AirSublimation}
From the coupled TGA-DSC measurement under dry air, the thermal and mass behaviour of the WO$_{3}$ powder was characterised with a series of thermal hold sections.

\begin{figure}[H]
     \centering
         \includegraphics[width=0.85\textwidth]{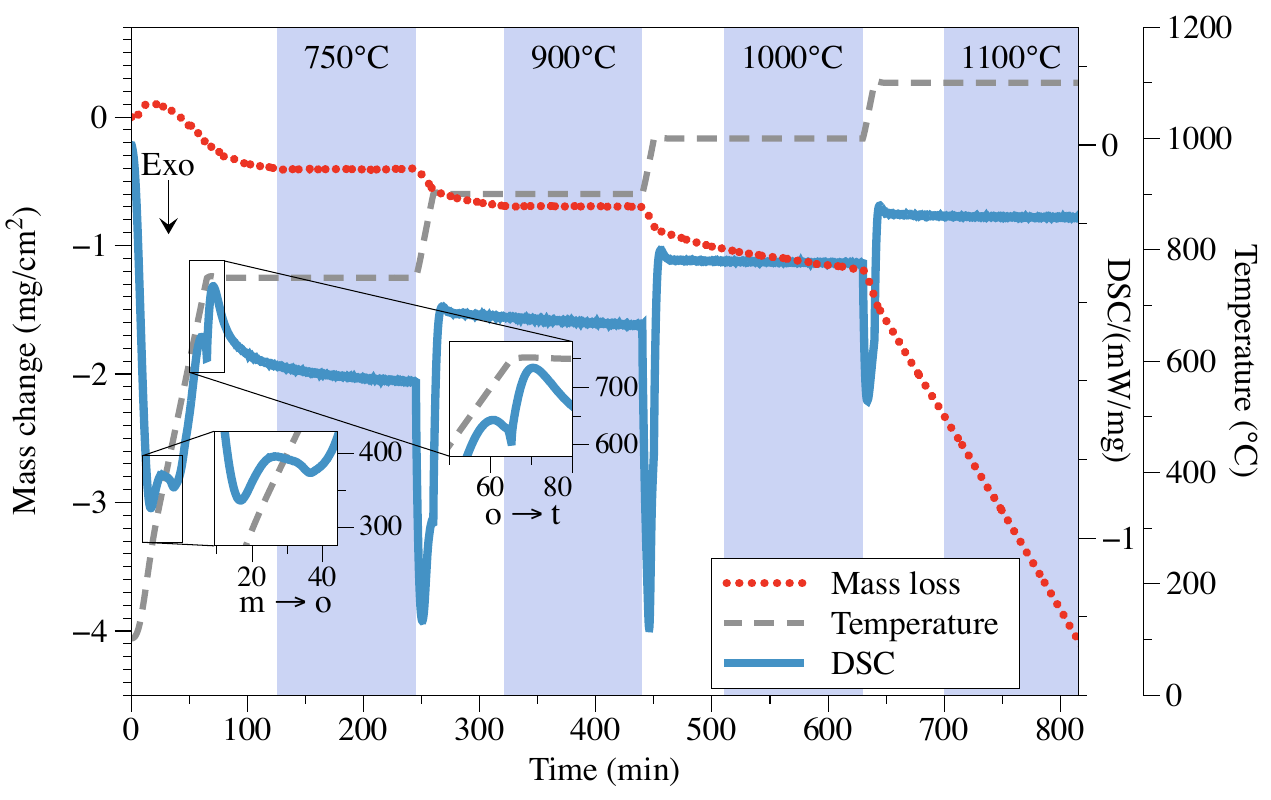}
        \caption{Mass loss and DSC curves of sublimation at temperatures between 750°C and 1100°C under synthetic air. The sublimation rate was calculated based on the average mass loss within the blue-shaded region. The two insets illustrate the WO$_{3}$ polymorphic transitions: monoclinic to orthorhombic (m → o) and orthorhombic to tetragonal (o → t), where m, o, and t denote the monoclinic, orthorhombic, and tetragonal phases, respectively.}
        \label{fig:Sub M&T&DSC}
\end{figure}

A correlation between exothermic responses and mass loss is shown in Fig. \ref{fig:Sub M&T&DSC}. Each heating segment was accompanied by a sharp exothermic peak and a rapid mass loss. This response can be attributed to two concurrent effects. First, as evidenced by the observations in Section \ref{Subsec:Materials}, the WO$_3$ powder used in this study contains sub-stoichiometric oxide species. During sublimation, these species can undergo in situ oxidation to form stoichiometric WO$_3$, a transformation known to be exothermic. Second, the use of an open (uncovered) crucible allows the vapor-phase WO$_3$ to escape freely, eliminating the compensating effect of vapor-phase heat capacity. As discussed by Tsioptsias et al. \cite{tsioptsias_2025_ExothermicContributions}, this can enhance the apparent endothermic shift, thereby amplifying the signal deviation associated with mass loss. Another factor influencing the DSC signal arises from the endothermic polymorphic transitions that occur during heating: from monoclinic to orthorhombic at 330~$^\circ$C and from orthorhombic to tetragonal at 740~$^\circ$C \cite{simchi_2014_StructuralOptical}. These appear as subtle upward inflections illustrated in the zoomed-in portion of Fig. \ref{fig:Sub M&T&DSC}. Once the set-point temperature is reached, the DSC signal stabilises to a quasi-horizontal baseline whose ordinate becomes increasingly endothermic as the temperature rises. Given that the heat capacity of WO$_3$ is essentially constant between 750~$^\circ$C and 1100~$^\circ$C \cite{han_2020_WO3Thermodynamic}, this trend is attributed to the increasing rate of sublimation at elevated temperatures.

For each isothermal hold, the TGA curve becomes linear after approximately one hour, indicating that a steady-state sublimation regime has been reached. The slope measured in the subsequent two hours (blue shading in Fig. \ref{fig:Sub M&T&DSC}) defines the sublimation rate. Between 750~$^\circ$C and 1100~$^\circ$C, the rate increases monotonically with temperature yet remains time-invariant within each temperature plateau, consistent with zero-order kinetics controlled by surface reaction.

\subsection{Microscopy analysis}
\label{Subsec: Microscopy analysis}

\begin{figure}[H]
     \centering
         \includegraphics[width=1\textwidth]{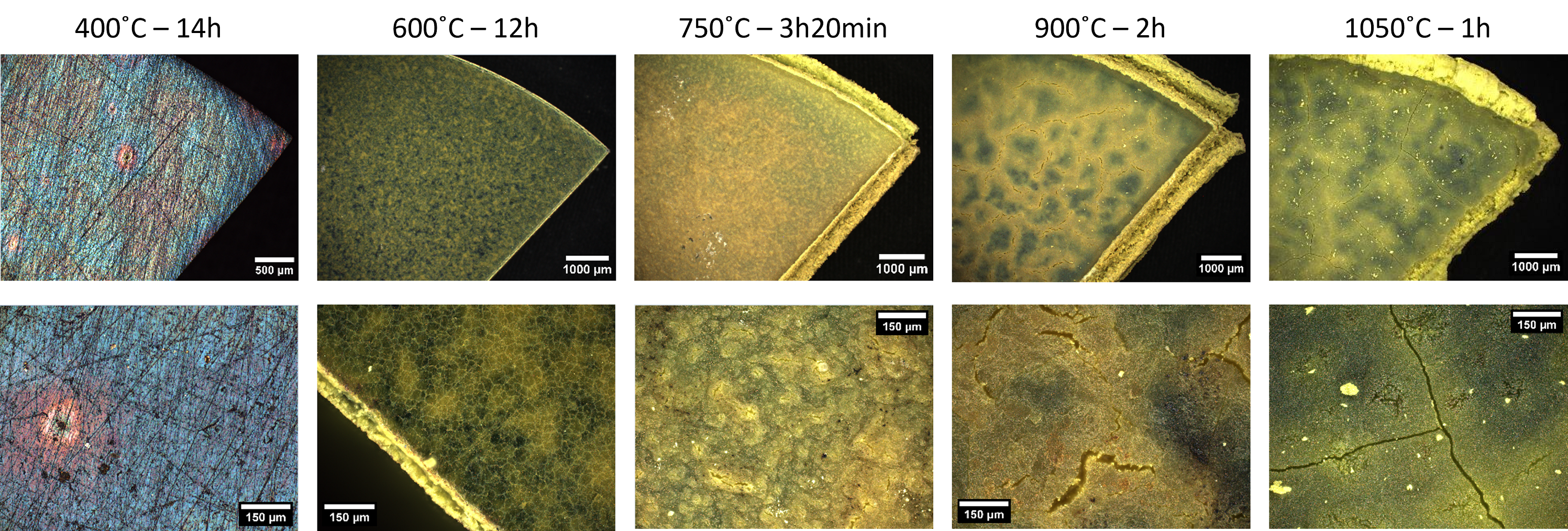}
        \caption{Visible light microscopy images of selected specimens after oxidation.}
        \label{fig:microscopy_overview}
\end{figure}

Representative visible light microscopy images of the surface of the W specimens after oxidation in dry air are displayed in Fig. \ref{fig:microscopy_overview}. The main features to be observed in these images are: (i) the presence of blue and yellow oxides, (ii) preferential oxide growth at the edge of the specimen, and (iii) cracks in the oxide scale.

\begin{figure}[H]
     \centering
         \includegraphics[width=1\textwidth]{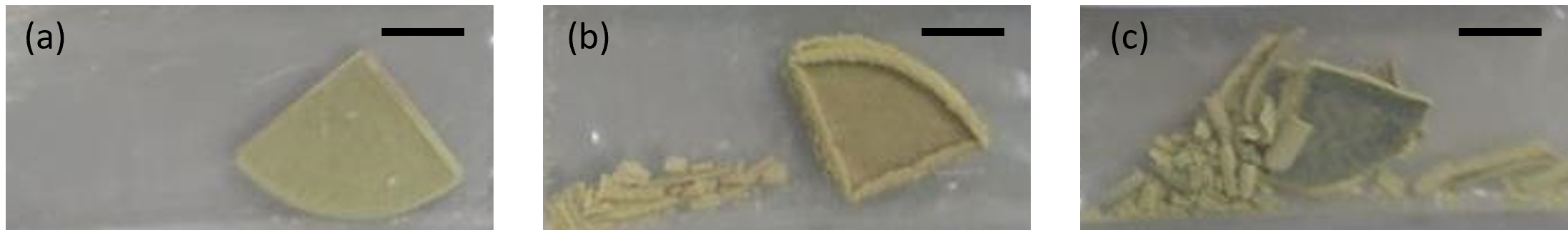}
        \caption{Photographs of selected specimens: (a) 10 hr at 700°C, (b) 10h at 850°C and (c) 1 hr at 1050°C. Photographs of specimens oxidised at 850°C and 1050°, show oxide scale that spalled off from the specimen during handling, next to the specimens. The scale bar is 5 mm.}
        \label{fig:photograph_oxide_spalling_edge}
\end{figure}

For thick oxide scale - where interference phenomena does not affect the colour - the colour of the bulk oxide can be associated with specific oxide scale phases: WO$_{2}$, WO$_{2.72}$, WO$_{2.9}$ and WO$_{3}$ are brown, violet, blue and yellow, respectively \cite{weil_beautiful_2013}. In Fig. \ref{fig:microscopy_overview}, the presence of blue and yellow areas in specimens oxidised at temperatures above 600°C indicates that the WO$_{2.9}$ (blue) phase is present alongside the WO$_{3}$ (yellow) phase. In the case of the specimen oxidised at 400°C for 14h, the oxide scale thickness is $<$ 1 \textmu m and it is expected that interference effects are significant. Therefore, it is difficult to use the colour to identify the oxide phase. Specimens oxidised at temperatures higher than 600°C show that the oxide is thicker at the edge than on the flat surface of the specimen. Fig. \ref{fig:photograph_oxide_spalling_edge} shows photographs of oxidised specimens with loose oxide parts, that spalled off from the edges when removing the specimen from the TGA instrument or when handling the specimen during post-oxidation characterisation. This indicates that the oxide scale at the specimen edge is very brittle and breaks off readily.

\subsection{X-ray diffraction}
\label{Subsec:XRD analysis}
\begin{figure}
    \centering
    \includegraphics[width=0.75\linewidth]{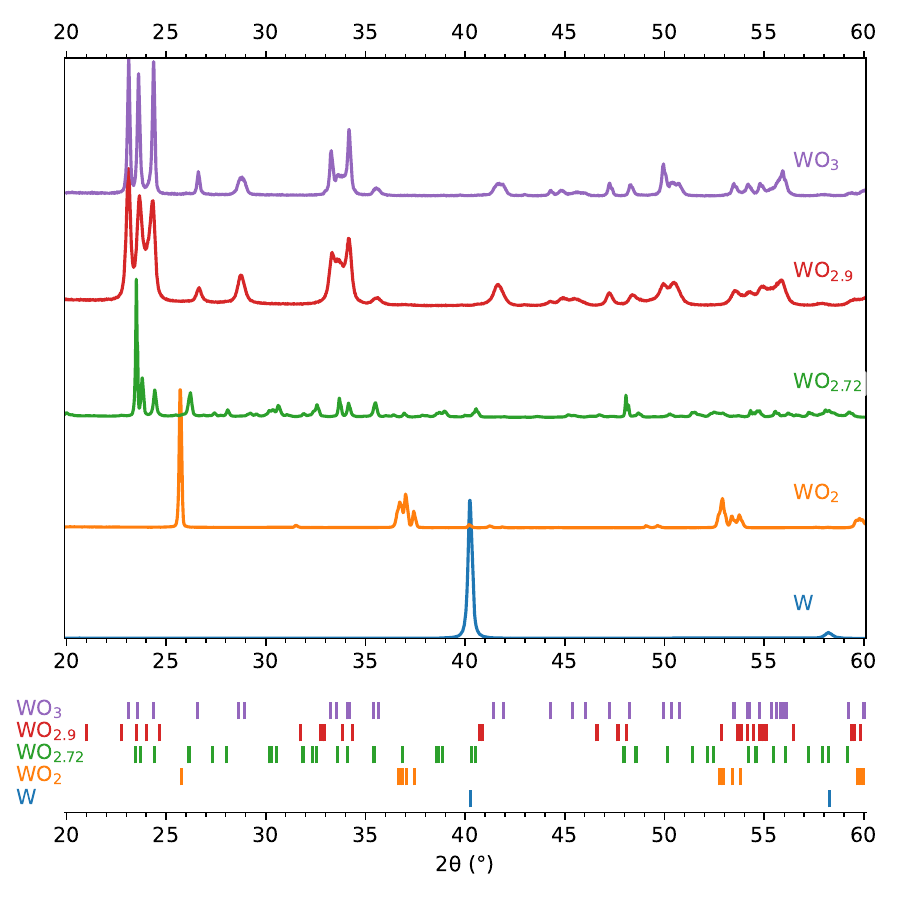}
    \caption{XRD patterns of pure W and reference WO$_{2}$, WO$_{2.72}$, WO$_{2.9}$ and WO$_{3}$ powders. The calculated positions of the diffraction peaks are displayed at the bottom of the figures for W and W oxide phases.}
    \label{fig:XRD_standards}
\end{figure}

X-ray diffraction (XRD) was used to identify the oxide phases formed in the oxide scale. Experimental XRD pattern of reference pure W and oxide powders were measured to be used as reference patterns for the interpretation of the XRD patterns of the oxidised specimens. Fig. \ref{fig:XRD_standards} shows these reference XRD patterns and their theoretical diffraction peak positions calculated from crystal structures retrieved from the Crystallography Open Database (COD) \cite{grazulis_crystallography_2009}. The following crystal structures were used: COD 1548686 (WO$_{2}$), COD 1538315 (WO$_{2.72}$), COD 1538317 (WO$_{2.9}$) and COD 2311041 (WO$_{3}$). For the WO$_{2}$, WO$_{2.72}$ and WO$_{3}$ phases, the XRD peak positions match with their theoretical values, while it doesn’t for the WO$_{2.9}$ phase. The best match of its XRD pattern is with the WO$_{3}$ phase – the peak positions are identical, but their peak intensities are different. This discrepancy can be explained by the distribution of the oxygen vacancies in the WO$_{2.9}$ phase, which is an oxygen-deficient WO$_{3}$ phase. The ordering of the oxygen vacancies required to form a Magneli phase (COD 1538317) \cite{kerr_relating_2025} is not observed experimentally, however the XRD pattern of the WO$_{2.9}$ powder could be explained by a random distribution of oxygen vacancies in the WO$_{3}$ phase. The analysis of reference oxide powders indicates that XRD is not suitable to distinguish between the WO$_{2.9}$ and WO$_{3}$ phases but can used to identify WO$_{2}$ and WO$_{2.72}$ phases, since they are distinctive from the WO$_{2.9}$/WO$_{3}$ oxide phases.

\begin{figure}[H]
     \centering
         \includegraphics[width=0.75\textwidth]{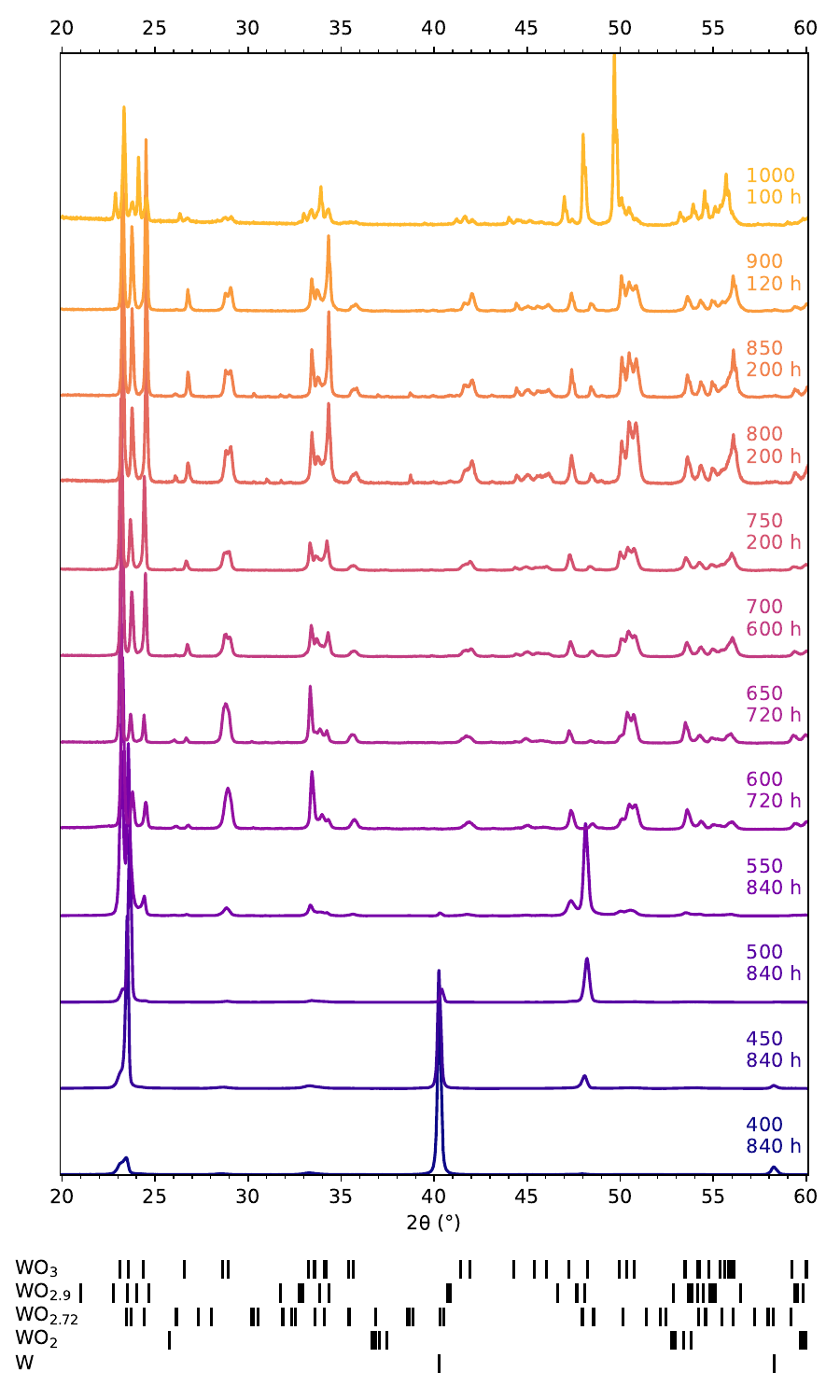}
        \caption{XRD patterns of specimens oxidised at temperatures between 400°C and 1000°C. The oxidation durations are displayed on the figure for each temperature. The calculated positions of the diffraction peaks are displayed at the bottom of the figures for W and W oxide phases.}
        \label{fig:XRD 400-1000C}
\end{figure}

\begin{figure}[H]
     \centering
     \begin{subfigure}{\textwidth}
        \centering
        \includegraphics[height=0.4\textheight]{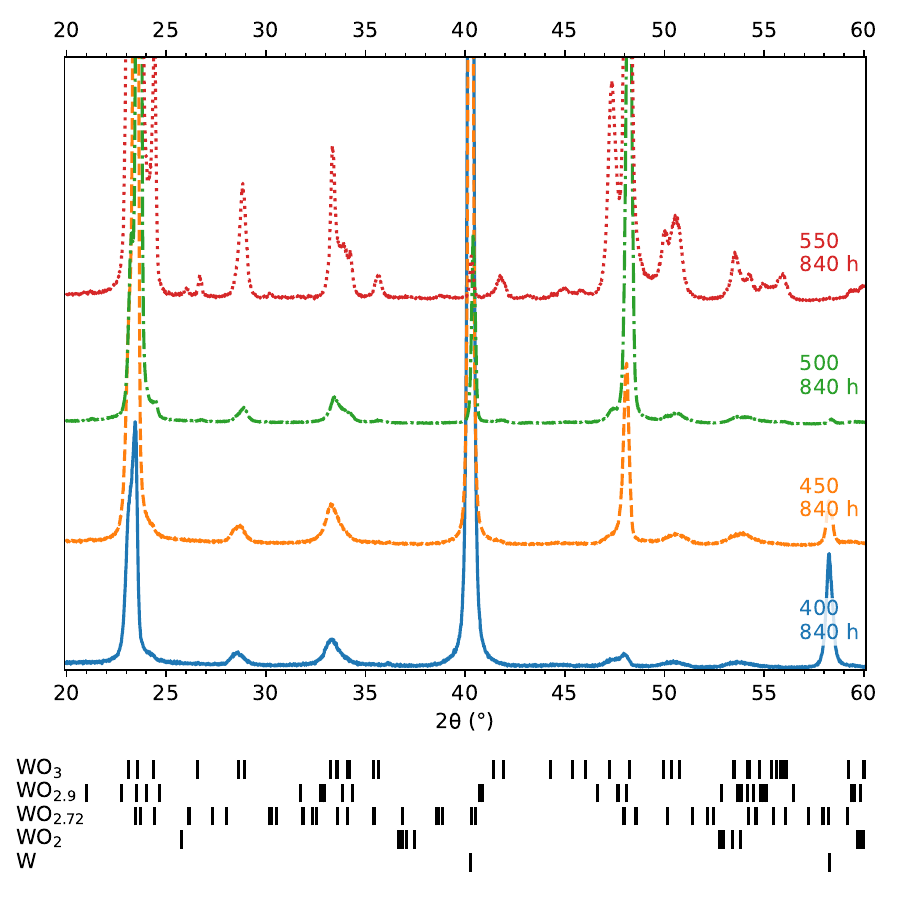}
        \caption{400°C to 550°C}
    \end{subfigure}
    \begin{subfigure}{\textwidth}
        \centering
        \includegraphics[height=0.4\textheight]{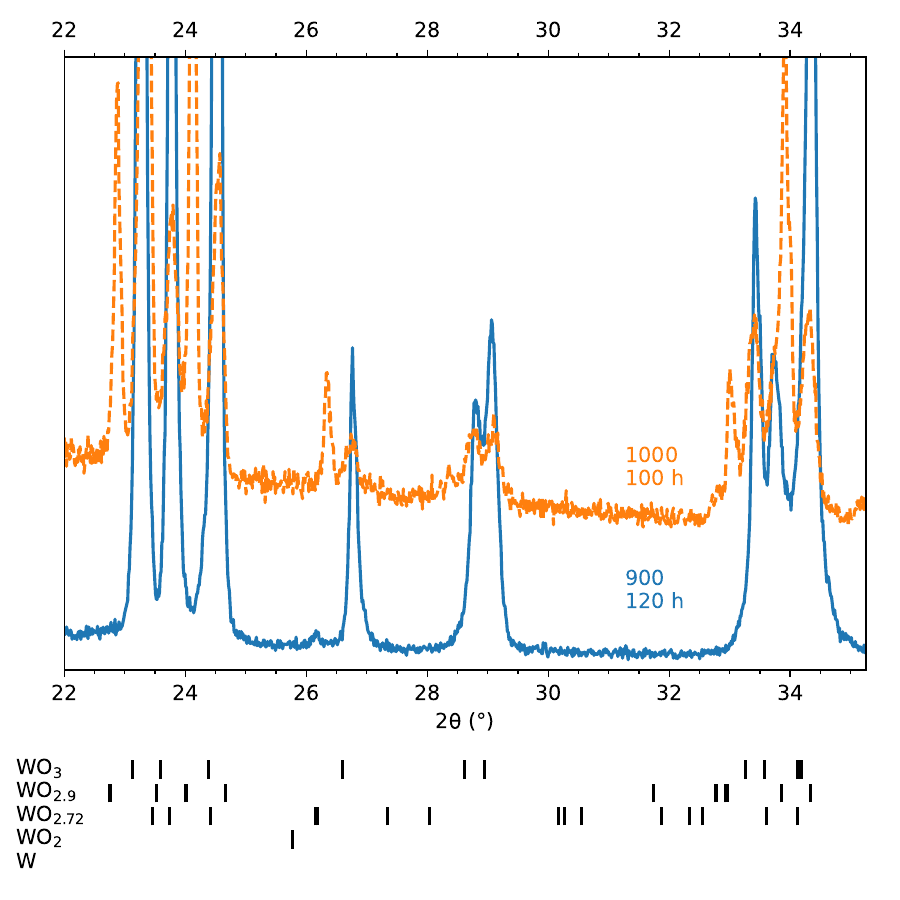}
        \caption{900°C and 1000°C}
    \end{subfigure}
    \caption{Zoom in of the XRD patterns of specimens oxidised at temperatures (a) between 400°C and 550°C and (b) 900°C and 1000°C. The oxidation durations are displayed on the figure for each temperature. The calculated positions of the diffraction peaks are displayed at the bottom of the figures for W and W oxide phases.}
    \label{fig:XRD zoom in}
\end{figure}

The XRD patterns of specimens oxidised between 400°C and 1000°C are displayed in Fig. \ref{fig:XRD 400-1000C}. At an oxidation temperature lower than 500 °C, the oxide scale is sufficiently thin ($\sim$ 5 \textmu m) to allow detection of the W substrate signal (peaks at 40° and 58° match the W crystal structure) while the peaks at 23° and 48° can be assigned to the WO$_{2.72}$, WO$_{2.9}$ and WO$_{3}$. The decrease of the contribution from the W crystal structure with increasing oxidation temperatures (and increasing oxide scale thicknesses) is consistent with the W signal coming from the W substrate. At higher oxidation temperatures (600°C to 900°C), the XRD patterns of oxidised specimens are in very good agreement with the reference pattern of WO$_{3}$ shown in Fig. \ref{fig:XRD_standards}. At an oxidation temperature of 1000°C, the XRD pattern shows variation in the peak intensities and additional peaks at 22.8°, 24.2° and 33°, as shown in Fig. \ref{fig:XRD zoom in}. These peaks can be assigned to the orthorhombic (Pbcn) polymorph of the WO$_{3}$ phases, which are formed at temperatures between 330°C and 740°C (see DSC measurement in Section \ref{Subsec:AirSublimation}) but have also been observed at room temperature after cooling down \cite{bruger_identification_2024}.

\subsection{Raman spectroscopy}
\label{Subsec: Raman spectroscopy}
\begin{figure}[H]
     \centering
        \includegraphics[width=1\textwidth]{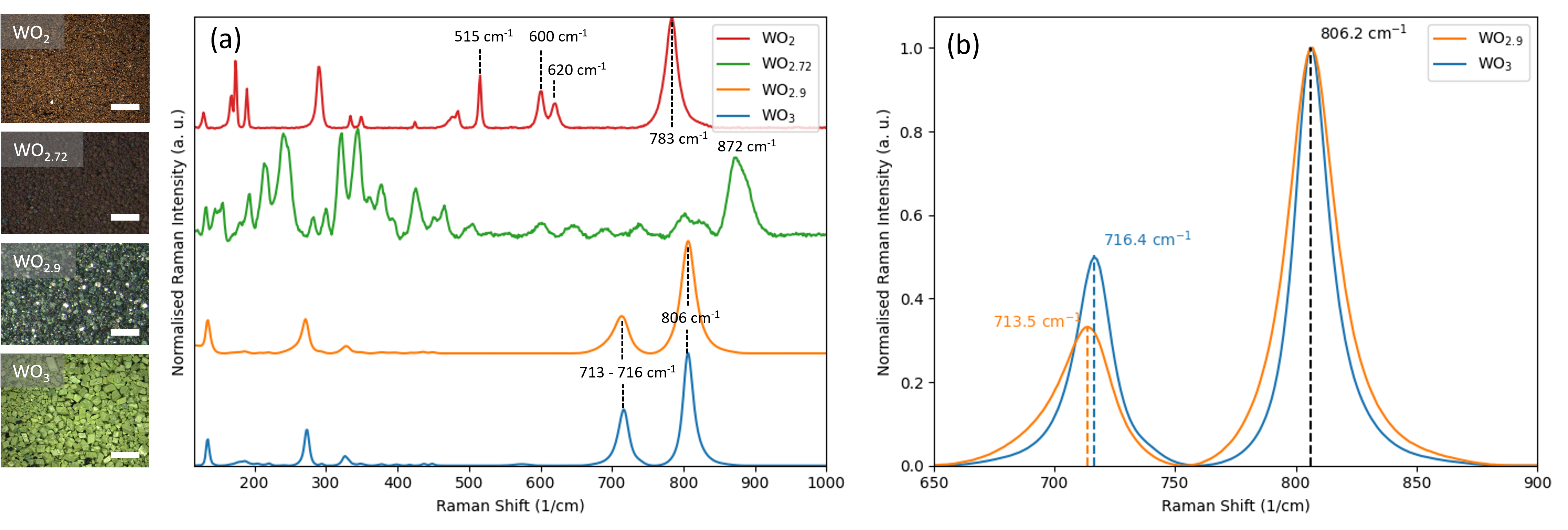}
        \caption{(a) Reference spectra of the four W oxide powders and their corresponding visible light microscopy images. The scale bar is 500 \textmu m. (b) Comparison of WO$_{2.9}$ and WO$_3$ powders in the 650-900 cm$^{-1}$ range showing the peak shift of the 716 cm$^{-1}$ peak of the latter to a lower wavelength in the former. The intensities have been normalised to their maximum values.}
        \label{fig:Raman_reference}
\end{figure}

Raman spectroscopy was performed to identify the W oxide phases formed and map the distribution of the various phases in plan view and cross section. To identify the oxide phases formed on the oxide scale in the metal samples, reference spectra of the oxide powders have been measured, and the spectra are shown in Fig. \ref{fig:Raman_reference}. The spectra of the WO$_{2.9}$ and WO$_{3}$ phases are very similar, while WO$_{2}$ and WO$_{2.72}$ have distinctive features (peak positions and widths) from the two other oxide phases. The WO$_{2.9}$ / WO$_{3}$ phases have strong peaks at 713-716 cm$^{-1}$ and 806 cm$^{-1}$, while the WO$_{2.72}$ phase has a characteristic peak at 872 cm$^{-1}$ and the WO$_{2}$ phase has peaks at 515 cm$^{-1}$, 600 cm$^{-1}$ and 620 cm$^{-1}$, which can be used to distinguish it from the others. Fig. \ref{fig:Raman_reference}(b) shows that the WO$_{2.9}$ and WO$_3$ phases can be distinguished by measuring the peak position in the range 713-716 cm$^{-1}$. These measurements agree with previous reports in the literature \cite{lu_raman_2007, gabrusenoks_infrared_2001, hijazi_tungsten_2017}. The comparison of the spectra of the WO$_{2}$ and WO$_3$ phases to the spectra reported in \cite{gabrusenoks_infrared_2001} shows that these spectra are characteristic of the monoclinic phase, which is the stable phase expected at room temperature. Since pure W only has acoustic phonons, it is not Raman active; i.e., no Raman signal will be measured from the pure W region. Peak positions and widths were fitted using Gaussian functions.

\begin{figure}[H]
     \centering
        \includegraphics[width=0.5\textwidth]{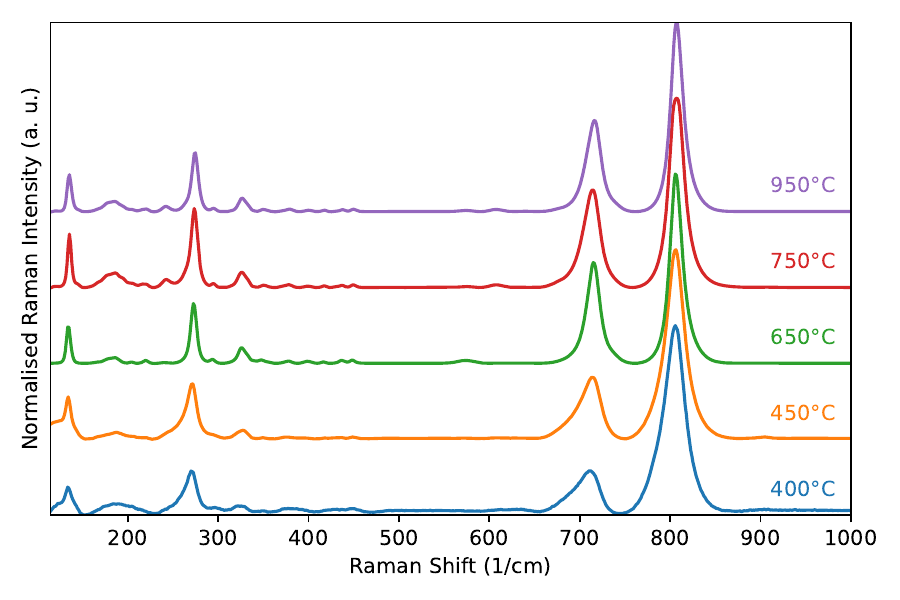}
        \caption{Raman spectra of specimens oxidised at temperatures of 400, 450, 650, 750 and 900°C. All these spectra have the peaks characteristic of the WO$_{2.9}$ or WO$_3$ phases as shown in Fig. \ref{fig:Raman_reference}.}
        \label{fig:Raman_plan_view}
\end{figure}

\begin{figure}[H]
     \centering
    \begin{subfigure}{0.33\textwidth}
        \centering
        \includegraphics[width=0.95\textwidth]{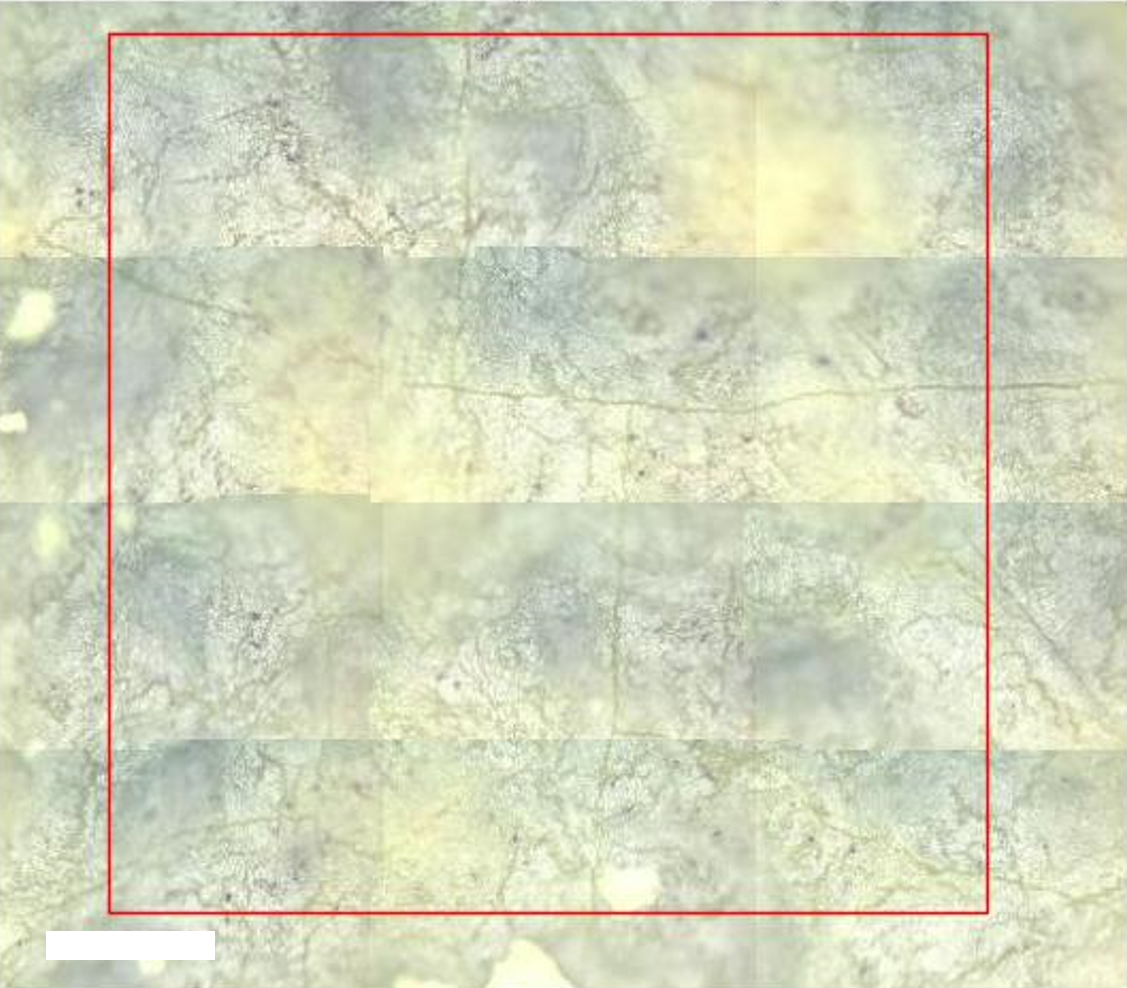}
        \caption{}
    \end{subfigure}\hfill
    \begin{subfigure}{0.33\textwidth}
        \centering
        \includegraphics[width=0.95\textwidth]{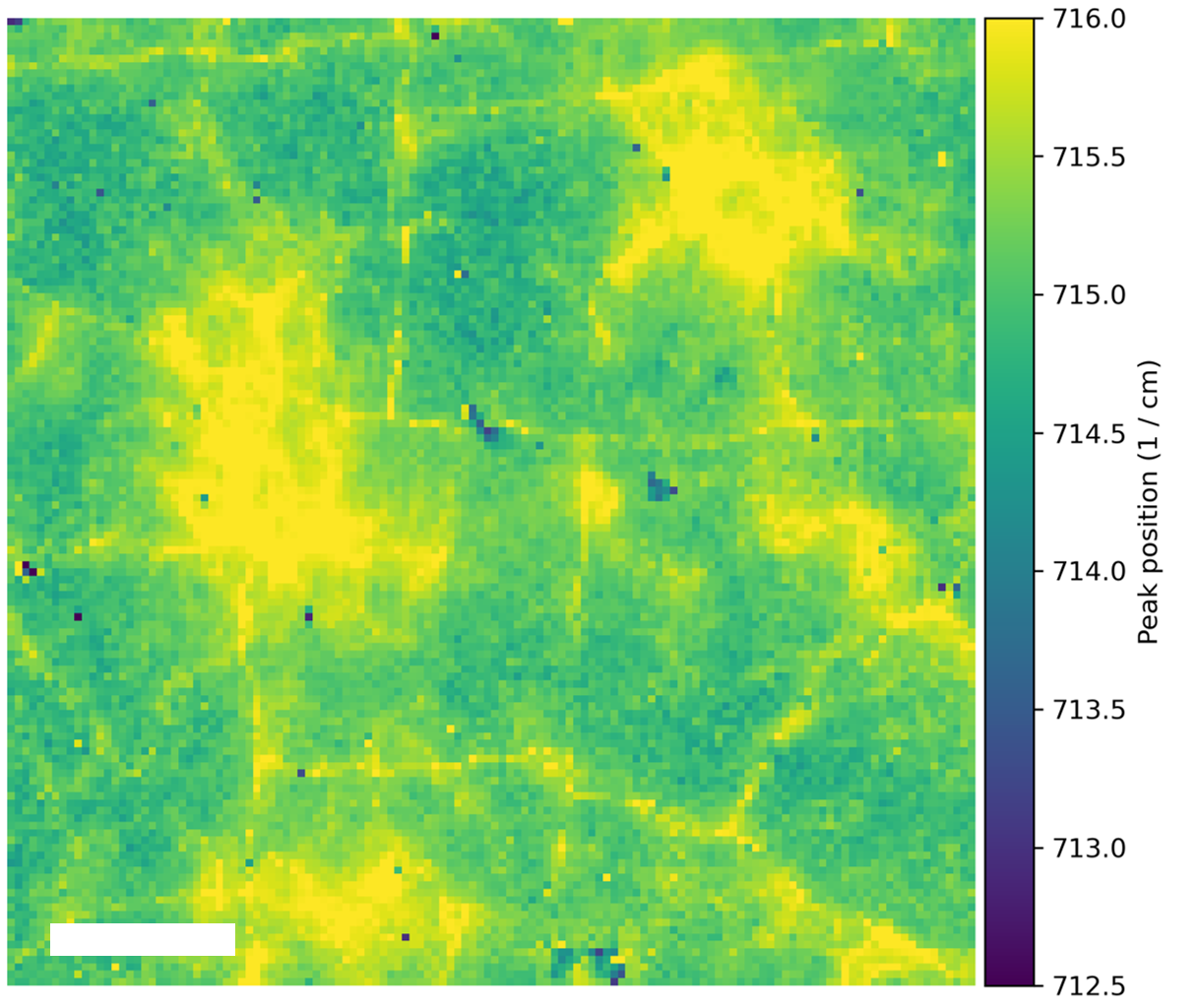}
        \caption{}
    \end{subfigure}
    \begin{subfigure}{0.33\textwidth}
        \centering
        \includegraphics[width=0.95\textwidth]{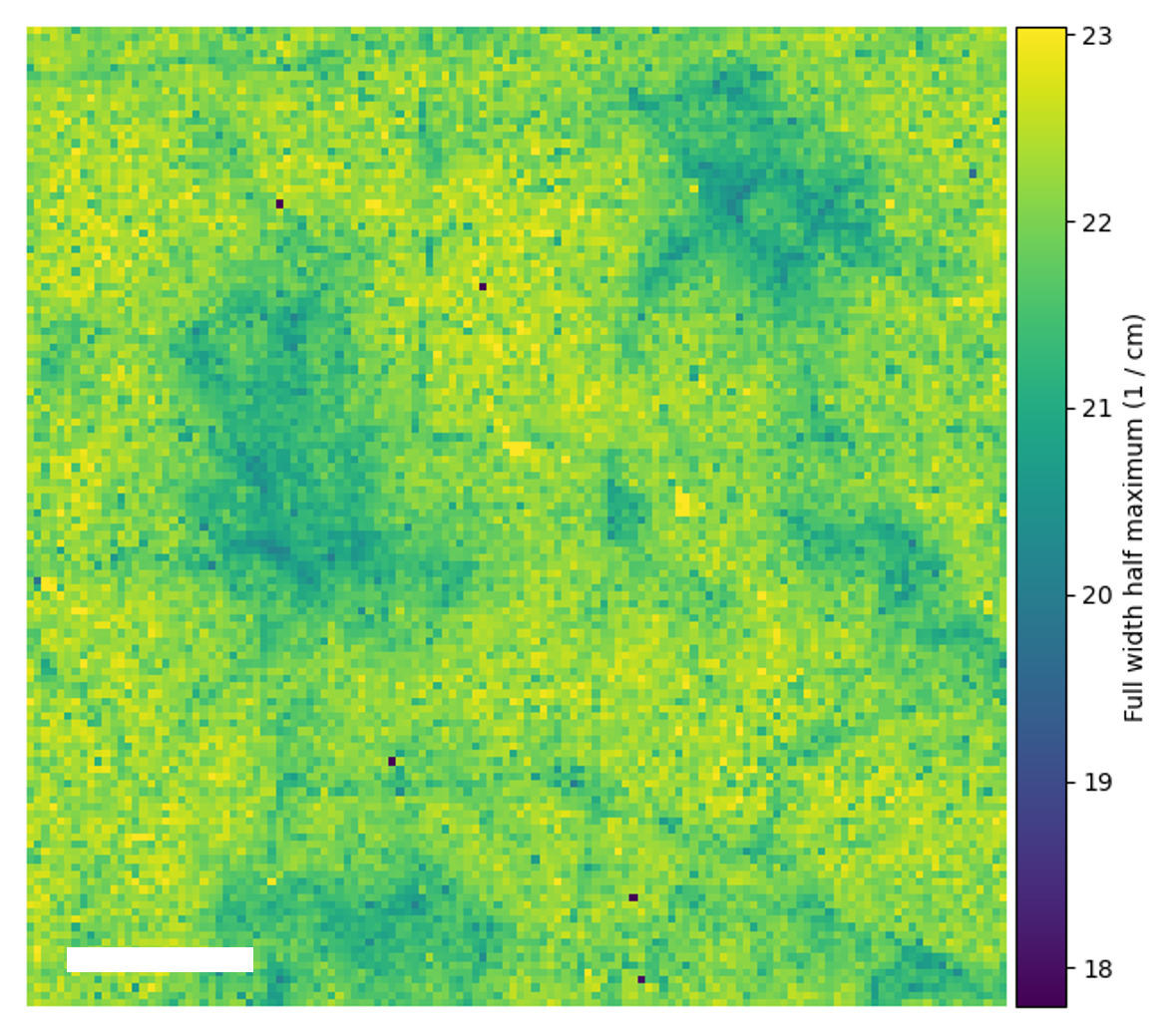}
        \caption{}
    \end{subfigure}
    \caption{Raman mapping of the WO$_{2.9}$ and WO$_3$ phases of a specimen oxidised at 950$^\circ$C for 2h. (a) Visible light microscopy image showing areas with blue and yellow oxides. (b-c) Peak position and full width half maximum of the peak at $\sim$715 cm$^{-1}$. The acquisition area is indicated by the red square in (a). The scale bar is 200 \textmu m.}
    \label{fig:Raman_plan_view_WO29_WO3}
\end{figure}

The Raman spectra measured in plan view of specimens oxidised at temperatures of 400, 450, 650, 750 and 900°C are displayed in Fig. \ref{fig:Raman_plan_view}. The comparison of these spectra with the reference spectra shown in Fig. \ref{fig:Raman_reference} indicates that the oxide phase formed at the surface layer corresponds to the WO$_{2.9}$ or WO$_3$ phases at all temperatures. As reported in Section \ref{Subsec: Microscopy analysis} (Fig. \ref{fig:microscopy_overview}), specimens oxidised at temperatures of about 500°C have a mixture of blue and yellow oxides, and the measurement of the peak at $\sim$715 cm$^{-1}$ allows identifying the presence of the WO$_{2.9}$ or WO$_3$ phases. The peak position and full width half maximum map displayed in Fig. \ref{fig:Raman_plan_view_WO29_WO3} confirms that the blue or yellow oxides at the surface of the oxidised specimen corresponds to the WO$_{2.9}$ or WO$_3$ phases, respectively.

\begin{figure}[H] 
     \centering
        \includegraphics[width=0.75\textwidth]{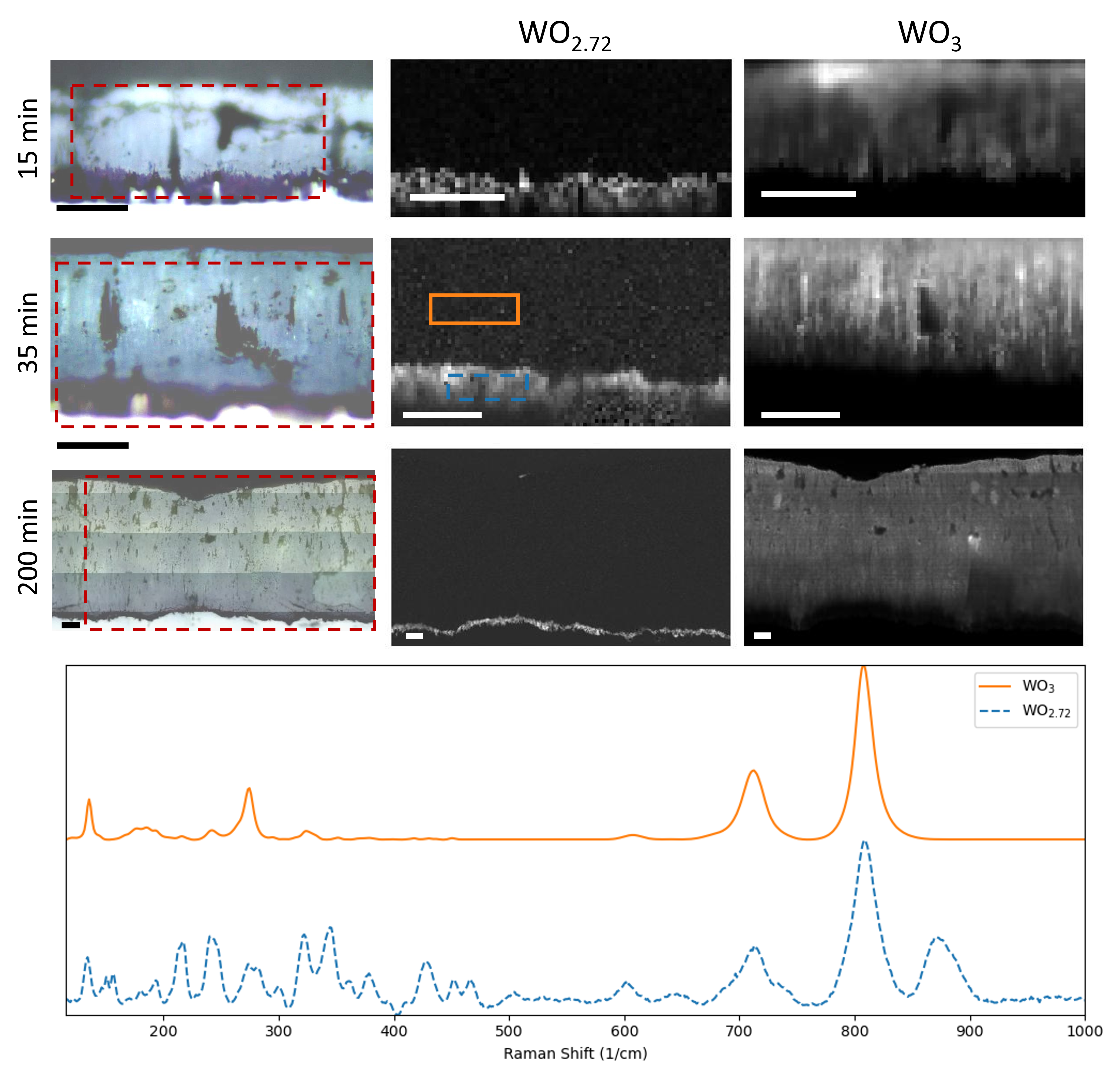}
        \caption{Cross-sectional Raman mapping of specimen oxidised at 850$^{\circ}$C for different oxidation times. The first column shows the visible light microscopy images, with the dashed line square showing where the Raman maps have been acquired. The second and last columns show the maps of the WO$_{2.72}$ and WO$_3$ phases, respectively. The bottom figure shows the Raman spectra acquired from the areas denoted by solid orange and blue lines in the WO$_{2.72}$ map at an oxidation time of 35 min. The scale bar is 20 \textmu m.}
        \label{fig:Raman_850C}
\end{figure}

Raman mapping was conducted on cross-sectioned specimens that had been oxidized at 850$^\circ$C for durations of 15, 35 and 200 min. The comparison of the spectra displayed in Fig. \ref{fig:Raman_850C} with the reference Raman spectra in Fig. \ref{fig:Raman_reference} shows the presence of two W oxide phases: a 10-20 \textmu m thin WO$_{2.72}$ layer between the W and the WO$_3$ layer. For longer oxidation times, the WO$_3$ layer increases in thickness, while the thickness of the WO$_{2.72}$ layer stays constant at 10-20 \textmu m. Raman mapping at other temperatures shows the same observation.

\begin{figure}[H] 
     \centering
        \includegraphics[width=0.75\textwidth]{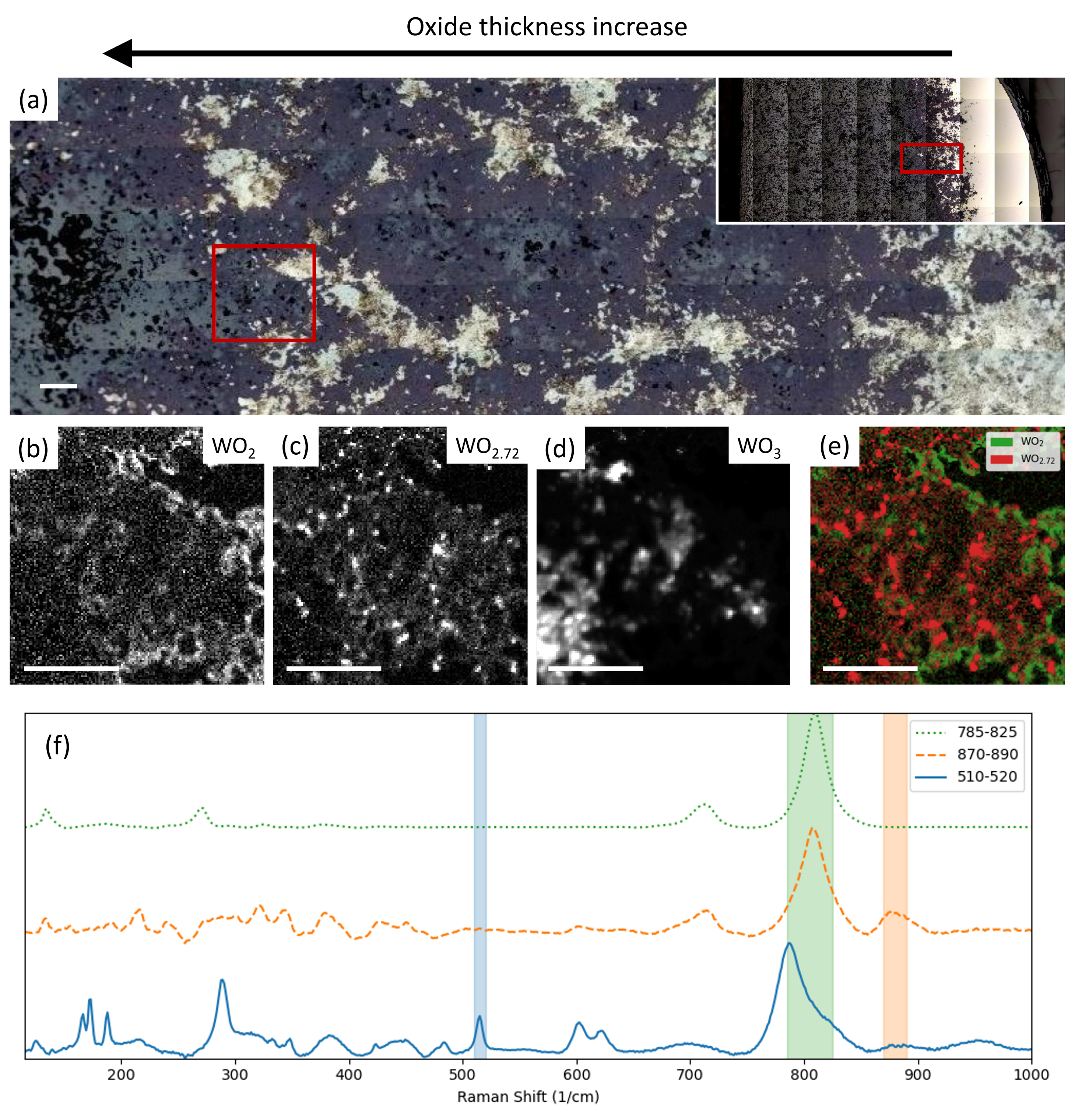}
        \caption{Raman mapping of 0.5$^{\circ}$ wedge polished specimen with oxide thickness increasing from right to left for a specimen oxidised at 750$^{\circ}$C for 200 min. (a) Visible light microscopy images showing full thickness oxide layer on the left-hand side and pure W on the right-hand side. The inset shows the full width of the specimen. (b-d) Raman maps of the WO$_2$, WO$_{2.72}$ and WO$_3$ phases, respectively. (e) Composite image of the WO$_2$, WO$_{2.72}$ phase maps. (f) Average Raman spectra of the WO$_2$, WO$_{2.72}$ and WO$_3$ phases maps with the shift range used to generate the maps. The scale bar is 50 \textmu m.}
        \label{fig:Raman_750C_wedge}
\end{figure}

In this study, the spatial resolution of Raman mapping is 1 \textmu m, which limits its ability to resolve oxide layers thinner than 1 \textmu m in cross-sectional analyses. An oxidised specimen polished in a wedge geometry has been prepared with a 0.5$^{\circ}$ polishing angle to allow measurement of oxide layers thinner than 1 \textmu m using Raman spectroscopy: for example a layer of 100 nm will appear as 11.5 \textmu m wide when projected at angle of 0.5$^{\circ}$ and will be above the spatial resolution of Raman mapping. Fig. \ref{fig:Raman_750C_wedge} shows the Raman mapping of a 0.5$^{\circ}$ wedge polished specimen oxidised at 750°C for 200 min, where the wedge direction is horizontal and the oxide thickness increases from right to left. In Fig. \ref{fig:Raman_750C_wedge} (a), pure W is shinny and is located in the right-hand side of the image, while the full thickness oxide appear as dark in the left-hand side. The comparison of the Raman spectra in Fig. \ref{fig:Raman_750C_wedge} (f) shows the presence of the WO$_2$, WO$_{2.72}$ and WO$_3$ phases. This identification is supported by the Raman maps in Fig. \ref{fig:Raman_750C_wedge} (b), (c), and (d), which were obtained by integrating the characteristic peaks: 510-520 cm$^{-1}$, 870-890 cm$^{-1}$ and 785-825 cm$^{-1}$ for WO$_2$, WO$_{2.72}$ and WO$_3$, respectively. The overlay of the WO$_2$, WO$_{2.72}$ phase maps in Fig. \ref{fig:Raman_750C_wedge} (e) shows that the intensity of WO$_2$ phases is highest at the interface between the pure W and the WO$_{2.72}$ phase. The high-intensity regions in the WO$_2$ map span a width of approximately 3-5 \textmu m, corresponding to an oxide scale thickness of 30-50 nm.

\subsection{Electron Microscopy}
\label{Subsec: Electron Microscopy analysis}

\begin{figure}[H]
     \centering
    \begin{subfigure}{0.5\textwidth}
         \centering
         \includegraphics[width=0.95\textwidth]{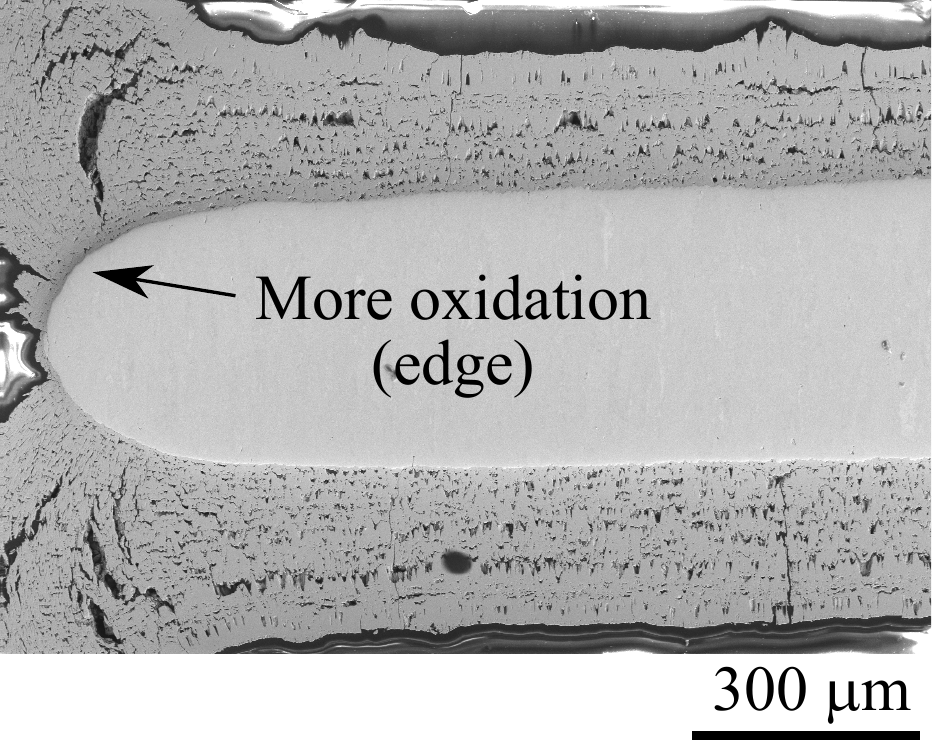}
            \caption{}
            \label{fig:CS_800C_edge}
         \end{subfigure}\hfill
    \begin{subfigure}{0.5\textwidth}
         \centering
         \includegraphics[width=0.95\textwidth]{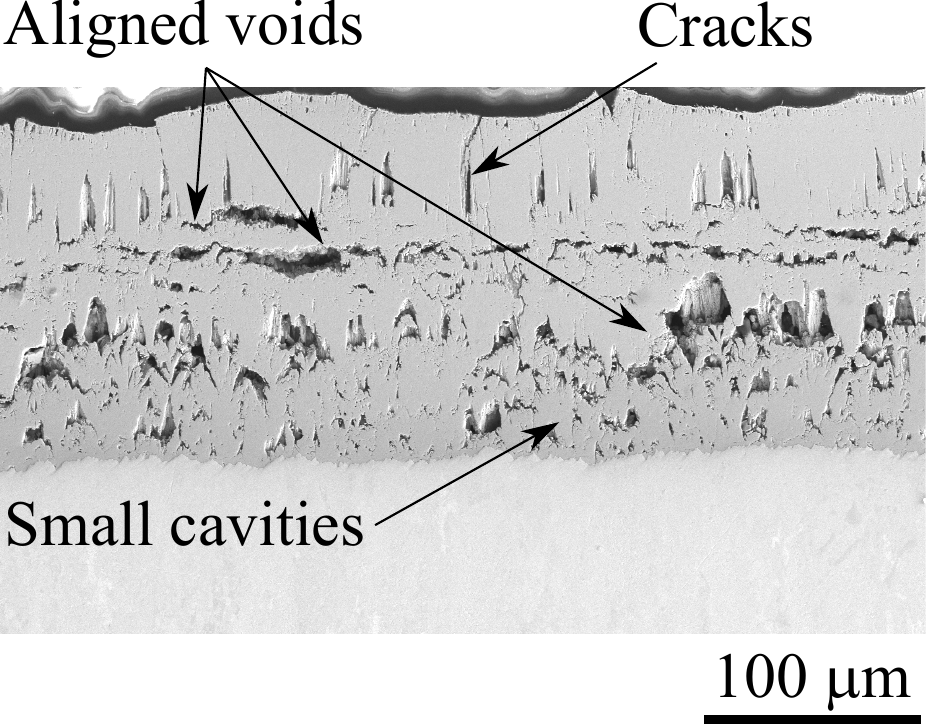}
            \caption{}
            \label{fig:CS_800C_center}
         \end{subfigure}\hfill
        \caption{SEM images of cross sections of a sample oxidised at 800°C for 200 minutes. (a) Edge; (b) Scales in the centre.}
        \label{fig:CS_800C}
\end{figure}

Scanning electron microscopy (SEM) was used to analyse the microstructural and morphological changes in the cross-sectional geometry.  Oxidation at 800°C for 200 minutes (Fig. \ref{fig:CS_800C} (\subref{fig:CS_800C_edge})) reveals a significant change in the shape of W, suggesting a more intense oxidation at the edges and noticeable delamination within the oxide layer (Fig. \ref{fig:CS_800C} (\subref{fig:CS_800C_center})) is observed. The outermost layer exhibits surface cracks but remains relatively dense. In contrast, the middle section contains three layers of aligned voids, with voids of the same orientation appearing at the same level. Numerous small, dense cavities are present closer to the W-oxide layer interface. The average thickness measured over 100 positions of the oxide layer was determined to be 207 \textmu m, which is in good agreement with the thickness of 225 \textmu m calculated from the mass gain measurement (Fig. \ref{fig:TGA curves} (\subref{fig:TGA curves linear})).

\section{Discussion}
\label{Sec:Discussion}

\subsection{Oxidation behaviour}
\label{Subsec:Kinetics}

Similarly to previous reports in the literature (see \cite{nagy2022oxidation} for a review), a transition from parabolic to linear kinetic regimes is observed in the oxidation rate of W. This suggests the existence of a bulk diffusion-controlled process in the early stage of oxidation, whereas the reaction is limited by surface processes after the transition to a linear regime. For each kinetic regime, the oxidation rates were calculated respectively by the following equations:
\begin{equation}
\left(\frac{\Delta m}{A} \right)^2=K_{p}t+C
\end{equation}
\begin{equation}
\frac{\Delta m}{A}=K_{l}t+C
\end{equation}
where $\Delta m$ is the mass change, $A$ is the surface area, $K_{p}$ and $K_l$ respectively denote the parabolic and linear rate constants, $t$ is the elapsed time, and $C$ is the integration constant.

\begin{figure}[H]
     \centering
    \begin{subfigure}{0.5\textwidth}
         \centering
         \includegraphics[width=\textwidth]{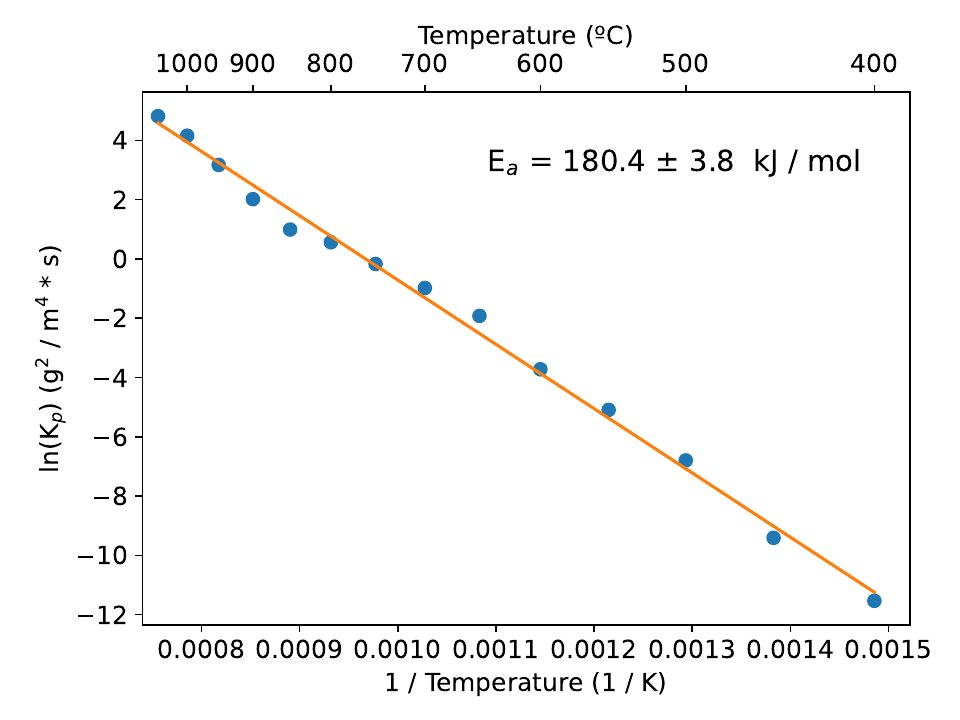}
            \caption{}
         \end{subfigure}\hfill
    \begin{subfigure}{0.5\textwidth}
         \centering
         \includegraphics[width=\textwidth]{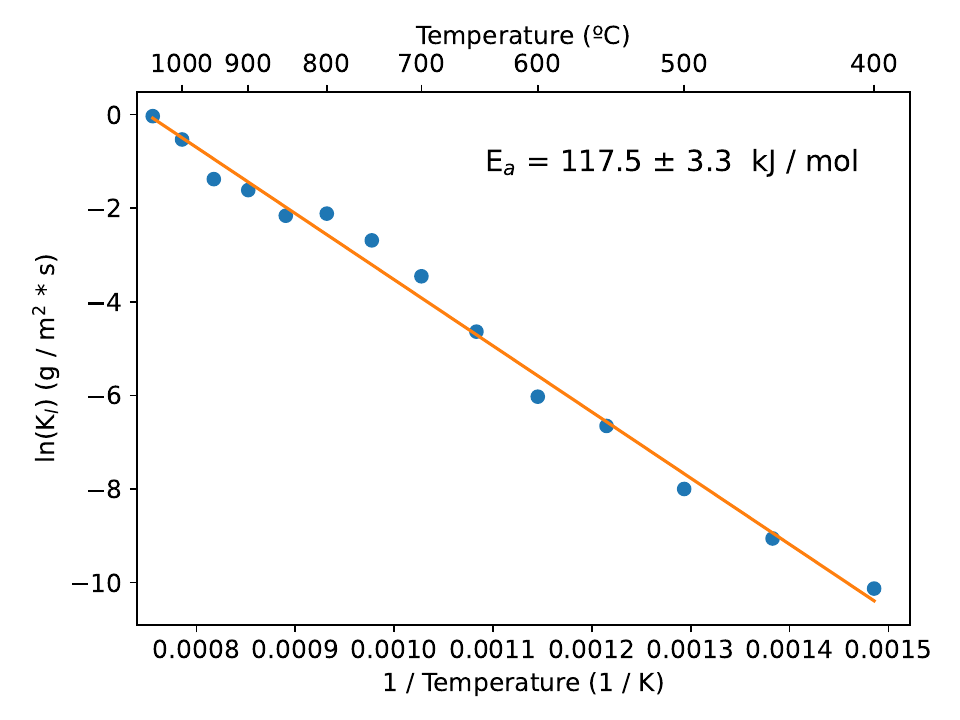}
            \caption{}
         \end{subfigure}\hfill
        \caption{Arrhenius plots of the (a) parabolic and (b) linear kinetic regimes of the oxidation rate. The corresponding activation energy $E_a$is displayed in each plot.}
        \label{fig:Activation energy}
\end{figure}

Fig. \ref{fig:Activation energy} shows the temperature dependence of the parabolic and linear oxidation rates and their corresponding fits with the Arrhenius equation:
\begin{equation}
\ K = K_0 \exp \left(-\frac{E_{a}}{R}\frac{1}{T} \right)
\label{arrhenius}
\end{equation}

\noindent where $K$ is the rate constant, $E_{a}$ is the activation energy, $K_{0}$ is the pre-exponential factor, $R$ is the gas constant, and $T$ is the absolute temperature in kelvin.

The activation energies of the parabolic and linear oxidation rates obtained from the fit are 180.4 $\pm$ 3.8 kJ/mol and 117.5 $\pm$ 3.3 kJ/mol, respectively. These values are in good agreement with values reported in the existing literature, with Nagy and Humphry-Baker reporting an activation energy of 204 kJ/mol and 118 kJ/mol for the parabolic and linear kinetics, respectively \cite{nagy2022oxidation}. Notably, the 900–1000°C data points for the parabolic segment were excluded from the fit by Nagy and Humphry-Baker; including them yields an activation energy of approximately 179 kJ/mol, which aligns closely with our results. Other discrepancies from those reported values are considered reasonable, given that Nagy and Humphry-Baker's data for the linear segment were derived from studies by five different authors. The discrepancy can be attributed to the variations in the experimental materials (specimen geometry), instrumentation, and conditions (oxidation duration and oxidising atmosphere) used in these studies. 

Fig. \ref{fig:microscopy_overview} and \ref{fig:CS_800C} show that preferential oxidation occurs at the edges of the specimen. It has been observed that the oxide at the edge of the specimens can spall easily from the surface during handling (Fig. \ref{fig:photograph_oxide_spalling_edge}). This could have an impact on the handling of W components and would need to be considered in the design of the maintenance and component handling procedures. It is observed that the measured oxidation rate increases with the perimeter-to-area ratio (PAR). Above a PAR value of ~2 (for circle and square geometries), it was estimated that the preferential oxidation at the edge is not negligible and that the oxidation rate may be underestimated. This may be relevant for W armors with design containing a large proportion of edges, such as castellation structures, where the dimension of the W block is 10x10x15 mm \cite{hong_preliminary_2014}, or micro-brush \cite{davis_assessment_1998, terra_icro-structured_2019}.

\begin{figure}[H]
     \centering
         \includegraphics[width=0.5\textwidth]{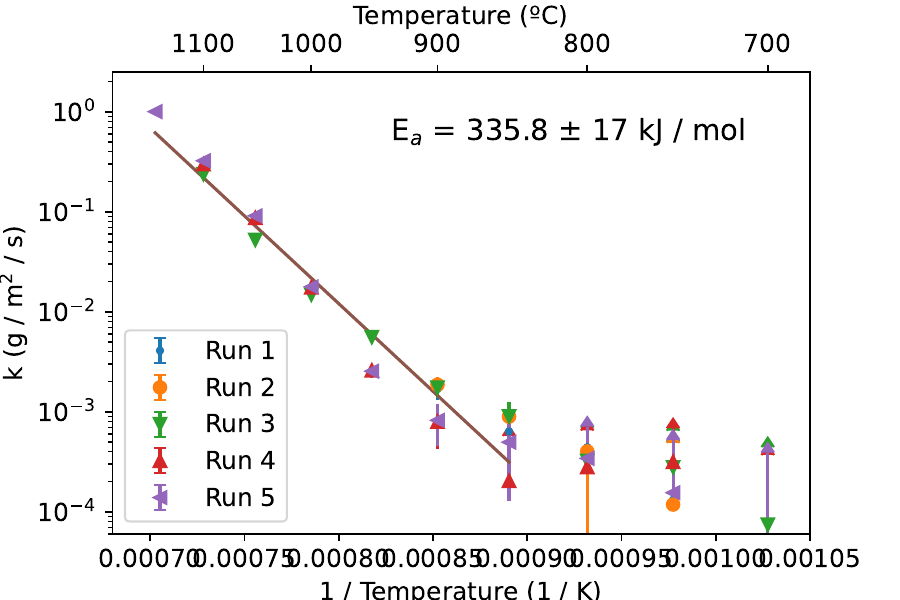}
        \caption{Arrhenius plot of the sublimation rate of WO$_3$ for five measurements using the same experiment conditions. Error bar with arrows indicates values below the estimated experimental error.}
        \label{fig:Ks}
\end{figure}

The sublimation rate of WO$_3$ is displayed in the Arrhenius plot in Fig. \ref{fig:Ks}. To assess reproducibility and estimate the sensitivity limit of the experimental setup, the measurements were reproduced 5 times and the sensitivity limit was estimated to be $3.7 \times 10^{-4}$ \textmu g / cm$^2$ /s. The activation energy of the sublimation rate is 335.8 $\pm$ 17 kJ/mol, which is lower than the value of 402.8 kJ/mol previously reported \cite{nagy2022oxidation}. This difference could be attributed to differences in experimental conditions, similarly to the case of oxidation rate as discussed above.

\subsection{Microstructure}
\label{Subsec:microstructure}

The XRD and Raman results presented in Sections \ref{Subsec:XRD analysis} and \ref{Subsec: Raman spectroscopy} show that the WO$_{2.9}$ and WO$_3$ phases are formed at the surface of the oxide scale for all temperatures investigated (above 400°C). According to Bandi and Srivastav \cite{bandi2021oxygen}, a low stoichiometric WO$_{2}$ layer with a thickness of only a few nanometers would be the first to be formed with the diffusion of oxygen. Initiated by WO$_{2}$, WO$_{2.72}$ is later formed and then gradually transformed into high stoichiometric WO$_{3}$. Even if no WO$_{2}$ could be detected using the XRD techniques throughout the whole characterisation of this work, the Raman results shown in Fig. \ref{fig:Raman_750C_wedge} confirmed that a thin 30-50 nm WO$_{2}$ layer is formed at the interface between the metal and the oxide scale. The observed layer thickness of 30-50 nm is consistent with the absence of WO$_{2}$ reflections in the XRD patterns, as such signals are likely lower than the background noise and therefore below the detection limit. Above the WO$_{2}$ layer, a 10-20 \textmu m thin WO$_{2.72}$ oxide layer is observed in wedge and cross-sectional geometry. The remaining oxide scale is formed of the WO$_{2.9}$ and WO$_3$ phases. Our experimental observations are consistent with the reports of Bandi and Srivastav \cite{bandi2021oxygen}. According to the interface tracking model developed by Huang et al. to simulate the kinetics evolution of W oxide scales, the time evolution of the oxide scales matches the parabolic growth when only one oxide sublayer such as WO$_{2.72}$ is involved \cite{huang2022multilayer}.

Our Raman results showing the presence of a thin WO$_2$ layer is relevant for detritiation of W. A common and efficient procedure to detritiate W is to anneal it in air at a temperature between 500 and 900°C \cite{stokes_detritiation_2023} to force the tritium to diffuse out of the material. In these conditions, W will oxidise and recent modelling results predict that the WO$_2$ phase is a tritium permeation barrier \cite{christensen_atomistic_2024}, which would mean that if this phase grows during detritiation it could negatively impact the detritiation efficiency. Our experimental results confirming the presence of this phase could explain the low detritiation efficiency of W observed in JET W tiles~\cite{stokes_detritiation_2023}.

\subsection{Parabolic to linear kinetics transition}

\begin{figure}[H]
   \centering
        \includegraphics[width=0.75\textwidth]{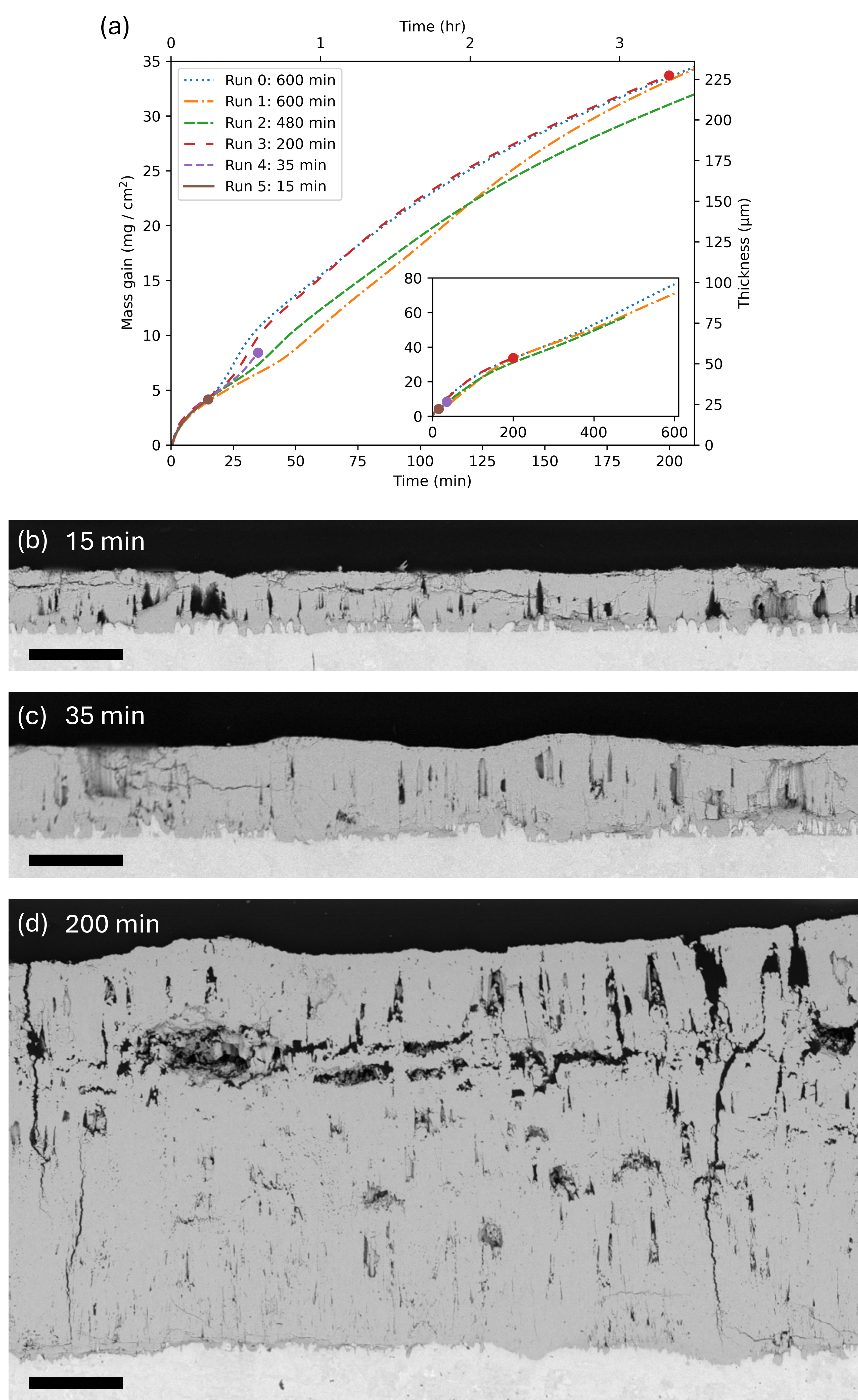}
       \caption{(a) Oxidation kinetics at 850°C and corresponding cross-section SEM images for (b) 15 min, (c) 35 min and (d) 200 min oxidation duration. The inset in (a) shows the full range (0 to 600 min) of the mass gain curves of the main figure. The states shown in (b), (c), and (d) correspond to the brown, purple, and red dots marked in (a), respectively. The scale bar in (b-d) is 50 \textmu m.}
       \label{fig:interrupted_test}
\end{figure}

The transition from parabolic to linear kinetics in the oxidation of W has been previously explained  by cracking occurring in the oxide scale \cite{CIFUENTES2012114}: cracks will cause the metallic surface to be exposed to oxidising species, causing localised and accelerated parabolic growth of new oxides. As such events occur concurrently at multiple sites across the specimen, the overall oxidation behaviour appears linear on a macroscopic scale \cite{Birks2012, CIFUENTES2012114}. To assess this hypothesis, three specimens were cross-sectioned at three different points of the parabolic to linear transition: the first point is at the end of the parabolic (15 min), the second is at the beginning of the kinetics transition (35 min) and the last point is at the beginning of the linear kinetics (200 min). The oxidation curve of specimens and their corresponding cross-sectional SEM images are displayed in Fig. \ref{fig:interrupted_test}. Additional oxidation curves are presented to evaluate the variability of the kinetic transition, particularly at 850°C, where the oxidation behaviour exhibits an 'S-shaped' profile (Fig. \ref{fig:TGA curves}(a) and \cite{gulbransen1960kinetics}).
The presence of large vertical voids at the end of the parabolic kinetics (15 min oxidation time, Fig. \ref{fig:interrupted_test} (b)) indicates that the presence of voids or cracking is not the cause of the transition to linear. An alternative explanation is that a non-porous oxide layer stops growth, as it was initially reported by Loriers in the case of oxidation of Ce \cite{loriers_loi_1950}. Our experimental results show (see Fig. \ref{fig:Raman_850C}) that the first 10-20 \textmu m of the W oxide scale, corresponds to the WO$_{2.72}$ layer and that this layer is dense. Moreover, the thickness of this layer does not increase more than ~20 \textmu m.
The multiple repeats of oxidation measurements at 850°C shows that in the parabolic regime, the mass gain curves overlap perfectly, while in the transition and the linear regime, the repeated measurements differ slightly. This could be explained by the fact in the parabolic regime, the kinetics is controlled through the diffusion of oxygen through the dense WO$_{2.72}$ layer, while in the linear regime, the formation of voids or cracks will differ slightly between repeated measurements and therefore the growth of oxide will differ slightly.

\subsection{Crack formation}
\label{Subsec:Mechanism of crack formation}

Previous studies have demonstrated that W forms fragile oxide layers at elevated temperatures. Cifuentes et al. \cite{CIFUENTES2012114} investigated the oxidation of pure W in the 600-800°C range and observed transitions in the TGA curves, indicating that the oxide layer failed to provide effective protection from the early stages of oxidation. They also reported an iterative cracking-oxidation cycle to explain the recurring cavities observed across the oxide scale, and attributed it to the significant volume change during oxidation. Cracking due to oxide volume expansion is usually measured with the Pilling-Bedworth ratio (PBR) as shown in Table \ref{tab:PBR}. In the case of W oxides, the PBR continues to increase as the stoichiometric ratio of oxygen increases, ranging from about 2 to 3.4, suggesting greater susceptibility to stress-induced cracking. For vertical through-thickness cracks, a high PBR is often considered to be a primary contributing factor \cite{CIFUENTES2012114,Birks2012}. For transverse cracks parallel to the oxide surface, they are likely caused by the thermal stresses generated during cooling. As reported by Besozzi et al.~\cite{besozzi_2018_CoefficientThermal}, there is a mismatch in the thermal expansion coefficient between the bulk metal and oxide ($4.2 \times 10^{-6}\ \mathrm{K}^{-1}$ for polycrystalline W and up to $8.9 \times 10^{-6}\ \mathrm{K}^{-1}$ for WO$_3$).

\begin{table}
\centering
\caption{\label{tab:PBR} Pilling-Bedworth ratio (PBR) calculation for W oxides of WO$_{3}$, WO$_{2.9}$, WO$_{2.72}$ and WO$_{2}$.}
\begin{tabular}{cccc}
\hline
Oxide type & W$_{WO_{x}}$ (g/mol)  & $\rho_{WO_{x}}$ (g/cm$^{3}$) \cite{lassner1999tungsten} & PBR \\
\hline
WO$_{3}$ & 231.84 & 7.27 & 3.32\\
WO$_{2.9}$ (W$_{20}$O$_{58}$) & 230.24 & 7.16 & 3.35\\
WO$_{2.72}$ (W$_{18}$O$_{49}$) & 227.39 & 7.78 & 3.05\\
WO$_{2}$ & 215.84 & 10.82 & 2.08\\
\hline
\end{tabular}
\end{table}

\begin{figure}[H]
     \centering
    \begin{subfigure}{0.49\textwidth}
        \centering
        \includegraphics[width=\textwidth]{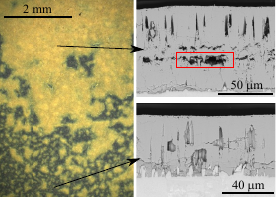}
        \caption{}
         \end{subfigure}\hfill
    \begin{subfigure}{0.495\textwidth}
        \centering
        \includegraphics[width=\textwidth]{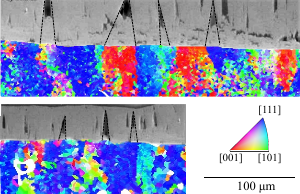}
        \caption{}
         \end{subfigure}\hfill
        \caption{Analysis of interrupted test at 850°C for 35min. (a) Image of oxidised surface and cross sections at the regions of thinner and thicker scale; (b) EBSD at the boundary between the raw material and the scale.}
        \label{fig:interrupted test}
\end{figure}

The surface morphology of an interrupted experiment of a specimen oxidised at 850$^{\circ}$C for 35 minutes reveals two main features of the oxide scale: heterogeneity in both colour and thickness. Analysis of the surface image obtained by visible light microscopy, together with cross-sectional SEM images (Fig. \ref{fig:interrupted test} (a)), reveals that the surface of the oxide scale comprises both blue planar regions and yellow oxide phases. The Raman spectroscopy results (Fig. \ref{fig:Raman_plan_view_WO29_WO3}) confirmed that the blue and yellow oxide are the WO$_{2.9}$ and WO$_3$ phases, respectively. Furthermore, the differently coloured regions also correspond to the variations in oxide scale thickness: blue regions indicate thinner scales while yellow regions indicate thicker ones. As highlighted within the red rectangle in Fig. \ref{fig:interrupted test} (a), the occurrence of transverse cracks is frequently associated with thicker yellow areas (i.e. WO$_{3}$). This suggests that differential thickness and transverse cracks are also associated with the type of oxide scale. Reported studies indicate that the unit cell volume undergoes a discontinuous expansion during the orthorhombic-to-tetragonal phase transition within the temperature range of 720$^{\circ}$C to 900$^{\circ}$C \cite{howard_2001_HightemperaturePhase}. In contrast, the Magnéli WO$_{2.9}$ phase exhibits no such phase transitions and therefore experiences no comparable crystallographic volume change \cite{cura_2021_MechanicalTribological}, resulting in fewer crack formations and lower thickness.

A further contributing factor to the thickness variation and crack initiation comes from the significant influence of grain orientation on the oxidation rate, with the \{100\} crystal planes oxidising preferentially and the planes near \{111\} oxidising the slowest \cite{SCHLUETER2019102}. Fig. \ref{fig:interrupted test} (b) shows two EBSD maps of the base metal W near the oxidation scale, where the main grain directions are [001] and [111]. The dashed lines in Fig. \ref{fig:interrupted test} (b) indicate the extrapolation of the contours of the arrowhead-shaped cavities toward the M:MO interface. These trajectories are mostly aligned along the boundaries between adjacent grains with the surface orientations of (001) and (111). Such a spatial correlation suggests that arrowhead-shaped cavity formation is preferentially initiated at sites of pronounced oxidation-rate mismatch.

\section{Conclusions}
\label{Sec:ConcludingRemarks}

This study provides a comprehensive analysis of W oxidation behaviour across a wide temperature range, highlighting critical structural degradation mechanisms relevant to fusion reactor applications. The oxidation rate of W has been measured over the temperature range of 400 to 1050°C, and the sublimation rate of WO$_3$ powder has been measured in synthetic air. Non-protective oxide scales were observed to grow with linear kinetics at all investigated temperatures. Above 600°C, the formation of voids and cracks has been observed, leading to the formation of dust or oxide spalling-phenomena that could pose challenges for accident (combined LOCA LOVA), maintenance and waste-handling scenarios of a fusion power plant. Preferential oxidation at the edges of samples (and equivalently, for example, tiles) is expected to further exacerbate dust formation or oxide spalling issues when handling components during maintenance. Raman spectroscopy revealed that the oxidation scale consists of three main layers: an inner WO$_2$ layer ($\sim$30-50 nm thick), a middle WO$_{2.72}$ layer (10-20 \textmu m thick), and an outer layer composed of WO$_{2.9}$/WO$_3$ phases. The presence of the WO$_2$ layer has been suggested as a possible explanation for the low detritiation efficiency observed during the detritiation of JET W tile \cite{stokes_detritiation_2023}.

\section{CRediT authorship contribution statement}

\textbf{Rongrui Li}: Investigation, Data Curation, Formal analysis, Writing - Original Draft. \textbf{Guillerno \'{A}lvarez}: Investigation, Data Curation, Validation, Formal analysis, Writing - Original Draft, Supervision. \textbf{Ayla Ipakchi}: Investigation. \textbf{Livia Cupertino Malheiros}: Writing - Review \& Editing, Funding acquisition. \textbf{Mark R. Gilbert}: Writing - Review \& Editing, Funding acquisition. \textbf{Emilio Mart\'{\i}nez-Pa\~neda}: Writing - Review \& Editing, Supervision, Funding acquisition. \textbf{Eric Prestat}, Conceptualization, Investigation, Data Curation, Methodology, Software, Validation, Writing - Original Draft, Supervision, Project administration, Funding acquisition.

\section{Declaration of competing interest}

The authors declare that they have no known competing financial interests or personal relationships that could have appeared to influence the work reported in this paper.

\section{Acknowledgements}
\label{Sec:Acknowledgeoffunding}
G. \'{A}lvarez acknowledges Margarita Salas Postdoctoral contract (Ref.: MU-21-UP2021-030) funded by the University of Oviedo through the Next Generation European Union. E. Martínez-Pañeda was supported by a UKRI Future Leaders Fellowship (grant MR/V024124/1). The authors acknowledge Andy London for commenting on the manuscript and support from the EPSRC (grant EP/R010161/1). This work has been funded by STEP, a major technology and infrastructure programme led by UK Industrial Fusion Solutions Ltd (UKIFS), which aims to deliver the UK’s prototype fusion powerplant and a path to the commercial viability of fusion. This work has been part-funded by the EPSRC Energy Programme [grant number EP/W006839/1] .

% \appendix

\small
\begin{thebibliography}{10}
\expandafter\ifx\csname url\endcsname\relax
  \def\url#1{\texttt{#1}}\fi
\expandafter\ifx\csname urlprefix\endcsname\relax\def\urlprefix{URL }\fi
\expandafter\ifx\csname href\endcsname\relax
  \def\href#1#2{#2} \def\path#1{#1}\fi

\bibitem{tokitani2019demonstration}
M.~Tokitani, S.~Masuzaki, T.~Murase, L.~E. Group, et~al., Demonstration of suppression of dust generation and partial reduction of the hydrogen retention by tungsten coated graphite divertor tiles in lhd, Nuclear Materials and Energy 18 (2019) 23--28.
\newblock \href {https://doi.org/10.1016/j.nme.2018.11.023} {\path{doi:10.1016/j.nme.2018.11.023}}.

\bibitem{koch2007self}
F.~Koch, H.~Bolt, Self passivating w-based alloys as plasma facing material for nuclear fusion, Physica Scripta 2007~(T128) (2007) 100.
\newblock \href {https://doi.org/10.1088/0031-8949/2007/T128/020} {\path{doi:10.1088/0031-8949/2007/T128/020}}.

\bibitem{zhang2023microstructure}
H.~Zhang, P.~R. Carriere, E.~D. Amoako, C.~D. Rock, S.~U. Thielk, C.~G. Fletcher, T.~J. Horn, Microstructure and elevated temperature flexure testing of tungsten produced by electron beam additive manufacturing, JOM 75~(10) (2023) 4094--4107.
\newblock \href {https://doi.org/10.1007/s11837-023-06045-5} {\path{doi:10.1007/s11837-023-06045-5}}.

\bibitem{HABAINY201826}
J.~Habainy, S.~Iyengar, K.~B. Surreddi, Y.~Lee, Y.~Dai, Formation of oxide layers on tungsten at low oxygen partial pressures, Journal of Nuclear Materials 506 (2018) 26--34.
\newblock \href {https://doi.org/10.1016/j.jnucmat.2017.12.018} {\path{doi:10.1016/j.jnucmat.2017.12.018}}.

\bibitem{klein_oxidation_2018}
F.~Klein, T.~Wegener, A.~Litnovsky, M.~Rasinski, X.~Y. Tan, J.~Gonzalez-Julian, J.~Schmitz, M.~Bram, J.~W. Coenen, C.~Linsmeier, Oxidation resistance of bulk plasma-facing tungsten alloys, Nuclear Materials and Energy 15 (2018) 226--231.
\newblock \href {https://doi.org/10.1016/j.nme.2018.05.003} {\path{doi:10.1016/j.nme.2018.05.003}}.

\bibitem{klein_studies_2020}
F.~Klein, Studies of oxidation resistant tungsten alloys at temperatures of 1100 {K} to 1475 {K}, {PhD}, Ruhr-Universität Bochum (Feb. 2020).
\newblock \href {https://doi.org/10.13154/294-7017} {\path{doi:10.13154/294-7017}}.

\bibitem{nagy2022oxidation}
D.~Nagy, S.~A. Humphry-Baker, An oxidation mechanism map for tungsten, Scripta Materialia 209 (2022) 114373.
\newblock \href {https://doi.org/10.1016/j.scriptamat.2021.114373} {\path{doi:10.1016/j.scriptamat.2021.114373}}.

\bibitem{noce_2021_NeutronicsAnalysisActivation}
S.~Noce, D.~Flammini, G.~Mariano, G.~Mazzone, F.~Moro, F.~Romanelli, R.~Villari, J.-H. You, Neutronics analysis and activation calculation for tungsten used in the {{DEMO}} divertor targets: {{A}} comparative study between the effects of {{WCLL}} and {{HCPB}} blanket, different {{W}} compositions and chromium, Fusion Engineering and Design 169  112428.
\newblock \href {https://doi.org/10.1016/j.fusengdes.2021.112428} {\path{doi:10.1016/j.fusengdes.2021.112428}}.

\bibitem{terentyev_2021_NeutronIrradiationHardening}
D.~Terentyev, C.~Yin, A.~Dubinko, C.~C. Chang, J.~H. You, Neutron irradiation hardening across {{ITER}} diverter tungsten armor, International Journal of Refractory Metals and Hard Materials 95  105437.
\newblock \href {https://doi.org/10.1016/j.ijrmhm.2020.105437} {\path{doi:10.1016/j.ijrmhm.2020.105437}}.

\bibitem{stokes_detritiation_2023}
T.~Stokes, M.~Damjanovic, J.~Berriman, S.~Reynolds, Detritiation of {JET} {Beryllium} and {Tungsten}, Fusion Science and Technology 0~(0) (2023) 1--7, publisher: Taylor \& Francis \_eprint: 10.1080/15361055.2023.2219826.
\newblock \href {https://doi.org/10.1080/15361055.2023.2219826} {\path{doi:10.1080/15361055.2023.2219826}}.

\bibitem{christensen_atomistic_2024}
M.~Christensen, E.~Wimmer, M.~R. Gilbert, C.~Geller, B.~Dron, D.~Nguyen-Manh, Atomistic modelling of tritium thermodynamics and kinetics in tungsten and its oxides, Nuclear Materials and Energy 38 (2024) 101611.
\newblock \href {https://doi.org/10.1016/j.nme.2024.101611} {\path{doi:10.1016/j.nme.2024.101611}}.

\bibitem{lassner1999tungsten}
E.~Lassner, W.-D. Schubert, Tungsten: properties, chemistry, technology of the elements, alloys, and chemical compounds, Springer Science \& Business Media, 1999.
\newblock \href {https://doi.org/10.1007/978-1-4615-4907-9} {\path{doi:10.1007/978-1-4615-4907-9}}.

\bibitem{gupta2003thermodynamics}
C.~K. Gupta, Thermodynamics and Kinetics, Chemical Metallurgy, 2003, pp. 225--342, 14; Wiley Online Books.
\newblock \href {https://doi.org/10.1002/3527602003.ch3} {\path{doi:10.1002/3527602003.ch3}}.

\bibitem{hauffe1995mechanism}
K.~Hauffe, The Mechanism of Oxidation of Metals—Theory, Oxidation of Metals, Springer US, Boston, MA, 1995, pp. 79--143, iD: Hauffe1995.
\newblock \href {https://doi.org/10.1007/978-1-4684-8920-0_3} {\path{doi:10.1007/978-1-4684-8920-0_3}}.

\bibitem{macdonald_1992_PointDefect}
D.~D. Macdonald, The {{Point Defect Model}} for the {{Passive State}}, Journal of The Electrochemical Society 139~(12)  3434.
\newblock \href {https://doi.org/10.1149/1.2069096} {\path{doi:10.1149/1.2069096}}.

\bibitem{sarrazin1996contribution}
P.~Sarrazin, A.~Galerie, M.~Caillet, Contribution to understanding parabolic oxidation kinetics of dilute alloys. part ii: Oxides with metal deficit or oxygen excess, Oxidation of Metals 46~(3) (1996) 299--312, iD: Sarrazin1996.
\newblock \href {https://doi.org/10.1007/BF01050801} {\path{doi:10.1007/BF01050801}}.

\bibitem{sikka1980oxidation}
V.~K. Sikka, C.~J. Rosa, The oxidation kinetics of tungsten and the determination of oxygen diffusion coefficient in tungsten trioxide, Corrosion Science 20~(11) (1980) 1201--1219, iD: 271639.
\newblock \href {https://doi.org/10.1016/0010-938X(80)90092-X} {\path{doi:10.1016/0010-938X(80)90092-X}}.

\bibitem{landolt2007corrosion}
D.~Landolt, Corrosion and surface chemistry of metal, EPFL Press, Lausanne, 2007.

\bibitem{kofstad1995defects}
P.~Kofstad, Defects and transport properties of metal oxides, Oxidation of Metals 44 (1995) 3--27.

\bibitem{Birks2012}
N.~Birks, G.~H. Meier, F.~S. F.~S. Pettit, Introduction to the high-temperature oxidation of metals, 2nd Edition, Cambridge University Press, Cambridge, UK, 2006.
\newblock \href {https://doi.org/10.1017/CBO9781139163903.002} {\path{doi:10.1017/CBO9781139163903.002}}.

\bibitem{samal2016high‐temperature}
S.~Samal, High‐Temperature Oxidation of Metals, High Temperature Corrosion, IntechOpen, Rijeka, 2016, p. Ch. 6.
\newblock \href {https://doi.org/10.5772/63000} {\path{doi:10.5772/63000}}.

\bibitem{gulbransen1960kinetics}
E.~Gulbransen, K.~Andrew, Kinetics of the oxidation of pure tungsten from 500 to 1300 c, Journal of the Electrochemical Society 107~(7) (1960) 619.
\newblock \href {https://doi.org/10.1149/1.2427787} {\path{doi:10.1149/1.2427787}}.

\bibitem{CIFUENTES2012114}
S.~Cifuentes, M.~Monge, P.~Pérez, On the oxidation mechanism of pure tungsten in the temperature range 600–800°c, Corrosion Science 57 (2012) 114--121.
\newblock \href {https://doi.org/10.1016/j.corsci.2011.12.027} {\path{doi:10.1016/j.corsci.2011.12.027}}.

\bibitem{mardare_2019_ReviewVersatility}
C.~C. Mardare, A.~W. Hassel, Review on the {{Versatility}} of {{Tungsten Oxide Coatings}}, physica status solidi (a) 216~(12)  1900047.
\newblock \href {https://doi.org/10.1002/pssa.201900047} {\path{doi:10.1002/pssa.201900047}}.

\bibitem{lunk2023molybdenum}
H.-J. Lunk, H.~Hartl, Molybdenum and tungsten: oxides, suboxides and oxide hydrates, ChemTexts 9~(2) (2023) 5.
\newblock \href {https://doi.org/10.1007/s40828-022-00175-0} {\path{doi:10.1007/s40828-022-00175-0}}.

\bibitem{SCHLUETER2019102}
K.~Schlueter, M.~Balden, Dependence of oxidation on the surface orientation of tungsten grains, International Journal of Refractory Metals and Hard Materials 79 (2019) 102--107.
\newblock \href {https://doi.org/10.1016/j.ijrmhm.2018.11.012} {\path{doi:10.1016/j.ijrmhm.2018.11.012}}.

\bibitem{hyperspy}
F.~de~la Peña, E.~Prestat, V.~T. Fauske, J.~Lähnemann, P.~Burdet, P.~Jokubauskas, T.~Furnival, C.~Francis, M.~Nord, T.~Ostasevicius, K.~E. MacArthur, D.~N. Johnstone, M.~Sarahan, J.~Taillon, T.~Aarholt, P.~Quinn, V.~Migunov, A.~Eljarrat, J.~Caron, T.~Nemoto, T.~Poon, S.~Mazzucco, actions user, N.~Tappy, N.~Cautaerts, S.~Somnath, T.~Slater, M.~Walls, pietsjoh, H.~Ramsden, Open source {Python} framework for exploring, visualizing and analyzing multi-dimensional data.
\newblock \href {https://doi.org/10.5281/zenodo.592838.} {\path{doi:10.5281/zenodo.592838.}}

\bibitem{pybaselines}
D.~Erb, {pybaselines}: A {Python} library of algorithms for the baseline correction of experimental data.
\newblock \href {https://doi.org/10.5281/zenodo.5608581} {\path{doi:10.5281/zenodo.5608581}}.

\bibitem{Bachmann2010}
F.~Bachmann, R.~Hielscher, H.~Schaeben, Texture analysis with mtex - free and open source software toolbox, Solid State Phenomena 160 (2010) 63--68.

\bibitem{panayotis_self-castellation_2017}
S.~Panayotis, T.~Hirai, V.~Barabash, A.~Durocher, F.~Escourbiac, J.~Linke, T.~Loewenhoff, M.~Merola, G.~Pintsuk, I.~Uytdenhouwen, M.~Wirtz, Self-castellation of tungsten monoblock under high heat flux loading and impact of material properties, Nuclear Materials and Energy 12 (2017) 200--204.
\newblock \href {https://doi.org/10.1016/j.nme.2016.10.025} {\path{doi:10.1016/j.nme.2016.10.025}}.

\bibitem{tsioptsias_2025_ExothermicContributions}
C.~Tsioptsias, Exothermic contributions in the {{DSC}} signal of endothermic effects, Journal of Thermal Analysis and Calorimetry\href {https://doi.org/10.1007/s10973-025-14145-4} {\path{doi:10.1007/s10973-025-14145-4}}.

\bibitem{simchi_2014_StructuralOptical}
H.~Simchi, B.~E. McCandless, T.~Meng, W.~N. Shafarman, Structural, optical, and surface properties of {{WO3}} thin films for solar cells, Journal of Alloys and Compounds 617  609--615.
\newblock \href {https://doi.org/10.1016/j.jallcom.2014.08.047} {\path{doi:10.1016/j.jallcom.2014.08.047}}.

\bibitem{han_2020_WO3Thermodynamic}
B.-y. Han, A.~V. Khoroshilov, A.~V. Tyurin, A.~E. Baranchikov, M.~I. Razumov, O.~S. Ivanova, K.~S. Gavrichev, V.~K. Ivanov, {{WO3}} thermodynamic properties at 80–1256~{{K}} revisited, Journal of Thermal Analysis and Calorimetry 142~(4)  1533--1543.
\newblock \href {https://doi.org/10.1007/s10973-020-09345-z} {\path{doi:10.1007/s10973-020-09345-z}}.

\bibitem{weil_beautiful_2013}
M.~Weil, W.~D. Schubert, \href{http://itia.info/assets/files/newsletters/Newsletter_2013_06.pdf}{The beautiful colours of tungsten oxides}, International tungsten Industry Association 4 (2013) 1--12.
\newline\urlprefix\url{http://itia.info/assets/files/newsletters/Newsletter_2013_06.pdf}

\bibitem{grazulis_crystallography_2009}
S.~Gražulis, D.~Chateigner, R.~T. Downs, A.~F.~T. Yokochi, M.~Quirós, L.~Lutterotti, E.~Manakova, J.~Butkus, P.~Moeck, A.~Le~Bail, Crystallography {Open} {Database} – an open-access collection of crystal structures, Journal of Applied Crystallography 42~(4) (2009) 726--729, number: 4.
\newblock \href {https://doi.org/10.1107/S0021889809016690} {\path{doi:10.1107/S0021889809016690}}.

\bibitem{kerr_relating_2025}
R.~D. Kerr, M.~R. Gilbert, S.~T. Murphy, Relating the formation energies for oxygen vacancy defects to the structural properties of tungsten oxides, Computational Materials Science 252 (2025) 113781.
\newblock \href {https://doi.org/10.1016/j.commatsci.2025.113781} {\path{doi:10.1016/j.commatsci.2025.113781}}.

\bibitem{bruger_identification_2024}
A.~Brüger, G.~Fafilek, M.~Neumann-Spallart, Identification of different {WO3} modifications in thin films for photocatalytic applications by peak shape analysis in high temperature {XRD} diffractometry, Journal of Photochemistry and Photobiology A: Chemistry 457 (2024) 115879.
\newblock \href {https://doi.org/10.1016/j.jphotochem.2024.115879} {\path{doi:10.1016/j.jphotochem.2024.115879}}.

\bibitem{lu_raman_2007}
D.~Y. Lu, J.~Chen, J.~Zhou, S.~Z. Deng, N.~S. Xu, J.~B. Xu, Raman spectroscopic study of oxidation and phase transition in {W18O49} nanowires, Journal of Raman Spectroscopy 38~(2) (2007) 176--180, \_eprint: https://onlinelibrary.wiley.com/doi/pdf/10.1002/jrs.1620.
\newblock \href {https://doi.org/10.1002/jrs.1620} {\path{doi:10.1002/jrs.1620}}.

\bibitem{gabrusenoks_infrared_2001}
J.~Gabrusenoks, A.~Veispals, A.~von Czarnowski, K.~H. Meiwes-Broer, Infrared and {Raman} spectroscopy of {WO3} and {CdWO4}, Electrochimica Acta 46~(13) (2001) 2229--2231.
\newblock \href {https://doi.org/10.1016/S0013-4686(01)00364-4} {\path{doi:10.1016/S0013-4686(01)00364-4}}.

\bibitem{hijazi_tungsten_2017}
H.~Hijazi, Y.~Addab, A.~Maan, J.~Duran, D.~Donovan, C.~Pardanaud, M.~Ibrahim, M.~Cabié, P.~Roubin, C.~Martin, Tungsten oxide thin film exposed to low energy {He} plasma: {Evidence} for a thermal enhancement of the erosion yield, Journal of Nuclear Materials 484 (2017) 91--97.
\newblock \href {https://doi.org/10.1016/j.jnucmat.2016.11.030} {\path{doi:10.1016/j.jnucmat.2016.11.030}}.

\bibitem{hong_preliminary_2014}
S.-H. Hong, E.-N. Bang, S.-T. Lim, J.-Y. Lee, S.~J. Yang, A.~Litnovsky, M.~Hellwig, D.~Matveev, M.~Komm, M.~van~den Berg, T.~Lho, C.~R. Park, G.-H. Kim, Preliminary test results on tungsten tile with castellation structures in {KSTAR}, Fusion Engineering and Design 89~(7) (2014) 1704--1708.
\newblock \href {https://doi.org/10.1016/j.fusengdes.2014.01.033} {\path{doi:10.1016/j.fusengdes.2014.01.033}}.

\bibitem{davis_assessment_1998}
J.~W. Davis, V.~R. Barabash, A.~Makhankov, L.~Plöchl, K.~T. Slattery, Assessment of tungsten for use in the {ITER} plasma facing components, Journal of Nuclear Materials 258-263 (1998) 308--312.
\newblock \href {https://doi.org/10.1016/S0022-3115(98)00285-2} {\path{doi:10.1016/S0022-3115(98)00285-2}}.

\bibitem{terra_icro-structured_2019}
A.~Terra, G.~Sergienko, M.~Tokar, D.~Borodin, T.~Dittmar, A.~Huber, A.~Kreter, Y.~Martynova, S.~Möller, M.~Rasiński, M.~Wirtz, T.~Loewenhoff, D.~Dorow-Gerspach, Y.~Yuan, S.~Brezinsek, B.~Unterberg, C.~Linsmeier, Micro-structured tungsten: an advanced plasma-facing material, Nuclear Materials and Energy 19 (2019) 7--12.
\newblock \href {https://doi.org/10.1016/j.nme.2019.02.007} {\path{doi:10.1016/j.nme.2019.02.007}}.

\bibitem{bandi2021oxygen}
S.~Bandi, A.~K. Srivastav, Oxygen-deficient tungsten oxides, Journal of Materials Science 56 (2021) 6615--6644.
\newblock \href {https://doi.org/10.1007/s10853-020-05757-2} {\path{doi:10.1007/s10853-020-05757-2}}.

\bibitem{huang2022multilayer}
S.~Huang, R.~Kerr, S.~Murphy, M.~R. Gilbert, J.~Marian, Multilayer interface tracking model of pure tungsten oxidation, Modelling and Simulation in Materials Science and Engineering 30~(8) (2022) 085015.
\newblock \href {https://doi.org/10.1088/1361-651X/aca111} {\path{doi:10.1088/1361-651X/aca111}}.

\bibitem{loriers_loi_1950}
J.~Loriers, \href{https://gallica.bnf.fr/ark:/12148/bpt6k3183z}{Loi d'oxydationdu cérium métallique. {Généralisation} à d'autres métaux}, Comptes rendus hebdomadaires des séances de l'Académie des sciences 231 (1950) 522.
\newline\urlprefix\url{https://gallica.bnf.fr/ark:/12148/bpt6k3183z}

\bibitem{besozzi_2018_CoefficientThermal}
E.~Besozzi, D.~Dellasega, A.~Pezzoli, A.~Mantegazza, M.~Passoni, M.~G. Beghi, Coefficient of thermal expansion of nanostructured tungsten based coatings assessed by substrate curvature method, Materials \& Design 137  192--203.
\newblock \href {https://doi.org/10.1016/j.matdes.2017.10.001} {\path{doi:10.1016/j.matdes.2017.10.001}}.

\bibitem{howard_2001_HightemperaturePhase}
C.~J. Howard, V.~Luca, K.~S. Knight, High-temperature phase transitions in tungsten trioxide - the last word?, Journal of Physics: Condensed Matter 14~(3)  377.
\newblock \href {https://doi.org/10.1088/0953-8984/14/3/308} {\path{doi:10.1088/0953-8984/14/3/308}}.

\bibitem{cura_2021_MechanicalTribological}
M.~E. Cura, M.~Trebala, Y.~Ge, P.~Klimczyk, S.-P. Hannula, Mechanical and tribological properties of {{WO2}}.9 and {{ZrO2}} + {{WO2}}.9 composites studied by nanoindentation and reciprocating wear tests, Wear 478--479  203920.
\newblock \href {https://doi.org/10.1016/j.wear.2021.203920} {\path{doi:10.1016/j.wear.2021.203920}}.

\end{thebibliography}


%% Text of bib
\end{document}